\documentclass[twocolumn,showpacs,preprintnumbers,prb,superscriptaddress]{revtex4}

\usepackage{graphicx}
\usepackage{dcolumn}
\usepackage{bm}

\begin{document}
\title{Spin-bias driven magnetization reversal and nondestructive detection in a single molecular magnet}
\author{Hai-Zhou Lu }
\affiliation{Department of Physics, and Centre of Theoretical and Computational Physics,
The University of Hong Kong, Pokfulam Road, Hong Kong, China}
\author{Bin Zhou }
\affiliation{Department of Physics, and Centre of Theoretical and Computational Physics,
The University of Hong Kong, Pokfulam Road, Hong Kong, China}
\affiliation{Department of Physics, Hubei University, Wuhan 430062, China}

\author{Shun-Qing Shen }
\affiliation{Department of Physics, and Centre of Theoretical and
Computational Physics, The University of Hong Kong, Pokfulam Road,
Hong Kong, China}
\date{\today }

\begin{abstract}
The magnetization reversal in a single molecular magnet (SMM) weakly
coupled to an electrode with spin-dependent splitting of chemical
potentials (spin bias) is theoretically investigated by means of the
rate equation. A microscopic mechanism for the reversal is
demonstrated by the avalanche dynamics at the reversal point. The
magnetization as a function of the spin bias shows hysteresis loops
tunable by the gate voltage and varying with temperature. The
nondestructive measurement to the orientation of giant spin in SMM
is presented by measuring the fully polarized electric current in
the response to a small spin bias. For Mn$_{12}$ac molecule, its
small transverse anisotropy only slightly violates the results
above. The situation when there is an angle between the easy axis of
the SMM and the spin quantization direction of the electrode is also
studied.
\end{abstract}

\pacs{75.50.Xx, 72.25.-b, 75.60.Jk, 72.25.Hg}
\maketitle


\section{Introduction}

Magnetization reversal driven by spin-polarized electric current had
attracted considerable interests over the last decade.
\cite{Berger1996prb549353,Slonczewski1996JMMM159.L1,Tsoi1998prl80.4281,Myers1999science285.867,Katine2000.prl.84.3149}
Recent experiments\cite{Kimura2006,Yang2008.NatPhys} demonstrated
that a pure spin current accompanied by no net charge current can
also be used to reverse the magnetization of a ferromagnetic
particle. In this setup, the particle is attached to a nonmagnetic
metal wire, in which the chemical potentials of two spin components
are split by using the non-local spin injection
technique.\cite{Johnson1985,Jedema2001,Jedema2002nat,Valenzuela2006}
By changing only the direction of injection current, the sign of the
splitting can be reversed, leading to the magnetization reversal in
both orientations.\cite{Yang2008.NatPhys} Application of only pure
spin current is appealing for spintronics devices, because it helps
to reduce critical currents, Joule heat, noise, etc.

Meanwhile, another family of intensively studied nanoscale magnetic
materials, the single molecular magnets
(SMM),\cite{Gatteschi2006book,Sessoli1993nature365141,Gatteschi1994science2651054,Friedman1996prl763830,
Thomas1996nature383145,Barra1997prb568192,Sangregorio1997prl784645,
Wernsdorfer1999science284133} was reported recently to be trapped in
a typical field effect transistor geometry, allowing electronic
transport measurement to be performed on an individual SMM with
great
tunability.\cite{Heersche2006prl96206801,Jo2006nl62014,Ni2006apl89212104,Henderson2007jap10109e102}
The experiment progresses inspired many transport theories of SMMs,
including magnetic signatures of SMMs in
transport,\cite{Kim2004prl92137203,Romeike2006prl96196805,Romeike2007prb75064404,Timm2007prb76014421}
Kondo
effect,\cite{Romeike2006prl96196601,Leuenberger2006prl97126601,Romeike2006prl97206601,Regueiro2007,Gonzalez2008prb78.054445}
Berry
phase,\cite{Leuenberger2006prl97126601,Gonzalez2007prl98256804} full
counting statistics,\cite{Imura2007prb75205341} quantum
computing,\cite{Lehmann2007nature2312}
cotunneling,\cite{Elste2007prb75195341} and vibrational
excitation.\cite{Regueiro2007}

Ion spins of magnetic metal in an SMM are interlocked to form a
collective giant spin, whose two maximally magnetized ground states
orient to opposite directions due to uniaxial anisotropy, a property
that implies to be a promising candidate for high-density
information storage. Therefore, one of the important issues is how
to manipulate and measure the magnetization of an
SMM,\cite{Romeike2006prl96196805,Timm2006prb73235304,Elste2006prb73235305,Misiorny2007epl7827003,Misiorny2007prb75134425,Misiorny2007prb76054448,Misiorny2008.prb.77.172414}
i.e., the processes of writing and reading qubit encoded by
SMM,\cite{Leuenberger2001nature410789,Zhou2002pra66.010301} using
transport approaches. It has been discussed that spin accumulation
can be induced by charge current.\cite{Romeike2007prb75064404}
Besides, it has been proposed that spin-polarized electric current
injected from ferromagnetic electrodes can be used to switch the
magnetization of
SMM.\cite{Timm2006prb73235304,Misiorny2007epl7827003,Misiorny2007prb76054448,Misiorny2008.prb.77.172414}

\begin{figure}[htbp]
\centering
\includegraphics[width=0.3\textwidth]{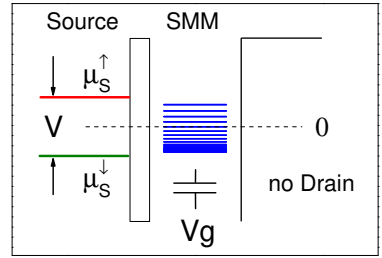}
\caption{Schematic of energy configurations of our setup to
manipulate the magnetization of an SMM. The splitting between
$\uparrow $ and $\downarrow $ Fermi levels of the only source lead
is phenomenologically denoted as $V$. Throughout the work the middle
point of $\mu_{\mathrm{S}}^{\uparrow /\downarrow }$ is set at 0 so
that $\mu_{\mathrm{S}}^{\uparrow /\downarrow }=\pm \frac{V}{2}$. The
horizontal lines in the SMM region correspond to resonant energies
to add an extra electron into the SMM via transitions from state
$|0,m\rangle$ to $|1,m\pm \frac{1}{2}\rangle^-$, which is tunable
with respect to $\mu_{\mathrm{S}}^{\uparrow/\downarrow}$ by the gate
voltage $V_g$.} \label{fig:model}
\end{figure}

Gold electrode is used in all the SMM transport experiments by
far,\cite{Heersche2006prl96206801,Jo2006nl62014,Ni2006apl89212104,Henderson2007jap10109e102}
which is also among the metals (Au, Ag, Al, Cu) employed in nonlocal
spin injection technique (for a review, see Ref.
\onlinecite{Valenzuela2007} and references therein). Therefore, it
is natural to expect that the magnetization reversal achieved in
ferromagnetic particle\cite{Kimura2006,Yang2008.NatPhys} could also
be realized in SMM, using the nonlocal spin injection. After all,
the magnetic moment of SMM is much smaller than that of
ferromagnetic particle. Although generating considerable
spin-dependent splitting of chemical potentials with high efficiency
remains a challenge, it has been demonstrated experimentally by many
other approaches such as the spin Hall
effect,\cite{Kato2004sci,Wunderlich2005} the spin pumping
effect\cite{Wang2006prl97.216602,Moriyama2006prl100.067602} and
incidence of polarized light into two-dimensional electron
gas.\cite{Cui2007,Li2006apl88.162105,Zhao2005.prb.72.201302,Zhao2006.prl.96.246601,Zhao2007.prb.75.075305}

Motivated by the progresses in both the non-local spin injection and
SMMs, in this work, we demonstrate theoretically that a pure spin
current induced by the spin-dependent splitting of chemical
potential (spin bias) in a nonmagnetic
electrode\cite{Wang2004.prb.69.205312,Sun2008.prb.77.195313,Lu2008prb77.235309,Xing2008.apl.93.142107,Stefanucci2008.prb.78.075425}
is enough to reverse the SMM magnetization, as shown in Fig.
\ref{fig:model}, where $V$ phenomenologically denotes
the spin dependent splitting of the Fermi levels for $\uparrow $ and $%
\downarrow $ electrons in the electrode, i.e., $\mu
_{\mathrm{S}}^{\uparrow /\downarrow }=\pm V/2$. We find that in the
context of the spin bias: (i) to reverse the SMM magnetization, only
one nonmagnetic electrode is needed. Neither magnetic
field\cite{Timm2006prb73235304} nor magnetic contact
\cite{Timm2006prb73235304,Misiorny2007epl7827003,Misiorny2007prb76054448}
is required. (ii) Only a pure spin current without accompanying a
net electric current flows in the process of reversal (Fig.
\ref{fig:avalanche}), which avoids the relaxation of magnetization
induced by electric current.\cite{Timm2006prb73235304} (iii) It
sheds a light on the mechanism of magnetization reversal from a
microscopic point of view, and may be extended to mesoscopic
magnetic particles or films.\cite{Kimura2006,Yang2008.NatPhys}

Moreover, we will discuss, in the context of using spin bias,
several effects not addressed or not clarified in the previous
literatures on the current-induced magnetic reversal in SMM: (i) By
analyzing the transition energy spectrum(Fig.
\ref{fig:transitionenergies}), we find that the activation energy at
which the magnetic reversal starts is determined not only by the
highest,\cite{Timm2006prb73235304,Misiorny2007prb76054448} but also
by the lowest transition energy, and is tunable by the gate voltage.
(ii) The SMM magnetization show magnetic hysteresis loop when
scanning the spin bias back and forth. The hysteresis loop can be
tuned by the gate voltage and shrinks with increasing temperature
(Fig. \ref{fig:hysteresis}). (iii) The avalanche dynamics at the
magnetic reversal point is demonstrated (Fig. \ref{fig:avalanche}),
which supports a microscopic magnetization reversal mechanism. (iv)
We show that the ground-state orientation of the giant spin in SMM
can be read out noninvasively, by measuring the charge current
through SMM driven by a small spin bias (Fig. {\ref{fig:measure}}).
(iv) The effect of weak transverse anisotropy is considered. (v) The
situation when there is an angle between the SMM easy axis and the
spin quantization direction of the electrode is discussed.

The paper is organized as follows. First, we will show that the
reversal mechanism is irrelevant to specific model used in Sec.
\ref{sec:mechanism}. The model and general formalism of theoretical
approach will be introduced in details in Sec. \ref{sec:model}. In
Sec, \ref{sec:numericals}, we present the numerical simulations of
the hysteresis loops tunable by the gate voltage, the avalanche
dynamics at the magnetic reversal point, and the nondestructive
detection to the orientation of the giant spin. In Sec.
\ref{sec:transverse}, the correction by the transverse anisotropy is
considered. In Sec. \ref{sec:theta}, the case when SMM easy axis is
not collinear with the spin quantization direction of the electrode
is investigated. Finally, a summary is presented to compare the
advantages of the present work to the existing proposals.

\section{\label{sec:mechanism}Model-irrelevant magnetization
reversal mechanism}
\begin{figure}[htbp]
\centering
\includegraphics[width=0.45\textwidth]{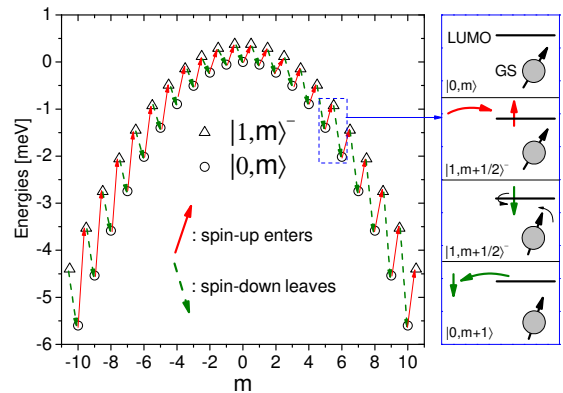}
\caption{Left: The two branches of molecular many-body states
considered in the simulations of this work. The basic parameters are
given in Sec. \ref{sec:parameters}. $V_g=-20$ mV. Arrows indicate
all the steps required to reverse the SMM magnetization from $-21/2$
to $21/2$. Right (four panels): The schematic of a single step that
increases the SMM magnetization by 1. This mechanism is supported by
the simulation shown in Figs. \ref{fig:avalanche} and
\ref{fig:avalanche probabilities}.} \label{fig:transitions}
\end{figure}

By far, many models are proposed to describe SMM, with extra
electrons added into it. These models include the giant spin
model,\cite{Heersche2006prl96206801,Jo2006nl62014,Romeike2006prl96196805,Timm2006prb73235304,Elste2006prb73235305,Misiorny2007prb76054448}
multi-ion model,\cite{Lehmann2007prl98117203} and those based on the
density-functional
theory.\cite{Park2004.prb.70.054414,Michalak.arxiv.0812.1058}

Despite the model employed, one can always select out two branches
of many-body states of SMM. One is for the neutral SMM, the other is
for when the SMM is charged with an extra electron. Assume the total
angular momentum of the ground state of the neutral SMM is $S$,
which has $2S+1$ states for different $z$ component of total angular
momentum, denoted as $|0,m\rangle$ ($m\in [-S,S]$). Adding the extra
electron is like coupling two angular momenta, leading to two
possible ground-state total angular momenta for the charged
branches, $S\pm \frac{1}{2}$, respectively. For simplicity, we
assume the total angular momentum of the ground state of the charged
branch is $S+\frac{1}{2}$, which has $2S+2$ states, denoted as
$|1,m\rangle$ ($m\in[-S-\frac{1}{2},S+\frac{1}{2}]$). Because of the
easy axis anisotropy, the ground states of the two branches are
$|0,\pm S\rangle $ and $|1,\pm(S+\frac{1}{2})\rangle$, respectively.
As an example, two such branches with uniaxial anisotropy barriers
are shown in Fig. \ref{fig:transitions}, using the giant spin model
proposed by Timm and
Elste.\cite{Timm2006prb73235304,Elste2006prb73235305}

Suppose one wants to reverse the giant spin originally orienting
antiparallel with the $z$ axis, i.e., at the state $|1,-21/2\rangle
$ or $|0,-10\rangle $, to parallel orientation, i.e., the state
$|1,21/2\rangle $ or $|0,10\rangle $. By connecting SMM to the lead,
one can generate a sequence of transitions that charge SMM with
spin-up electrons and discharge SMM with spin-down electrons
\begin{eqnarray}
&&|1,-21/2\rangle \rightarrow |0,-10\rangle \rightarrow
|1,-19/2\rangle\rightarrow ...\rightarrow |1,-11/2\rangle\nonumber\\
&&\rightarrow ...\rightarrow |0,0\rangle \rightarrow ...\rightarrow
|1,11/2\rangle\rightarrow ...\rightarrow |1,19/2\rangle\nonumber\\
&&\rightarrow |0,10\rangle \rightarrow |1,21/2 \rangle,
\end{eqnarray}
as shown in Fig. \ref{fig:transitions}. During each of these
transitions, the magnetic moment of an electron spin is transferred
to SMM, by either adding an spin-$\uparrow $ electron into or
removing a spin-$\downarrow $ electron from SMM. The right four
panels in the Fig. \ref{fig:transitions} depict a typical step, in
which an $\uparrow $ spin enters, flips to $\downarrow $ spin owing
to the exchange coupling while increases the giant spin orientation
by one unit, and escapes from SMM. Such a step repeats until the
giant spin orientation is reversed.

Energetically, to generate this charging-discharging sequence, the
Fermi level for the spin-up electrons in the lead should be higher
than all the transition energies of adding an spin-up electron from
a neutral state $|0,m\rangle$ to a charged state
$|1,m+\frac{1}{2}\rangle$, i.e.,
\begin{eqnarray}
\mu^{\uparrow}_{\mathrm{S}} >
E_{|1,m+\frac{1}{2}\rangle}-E_{|0,m\rangle},
\end{eqnarray}
while the spin-down Fermi level of the lead should be lower than all
the transition energies of adding an spin-down electron from a
neutral state $|0,m\rangle$ to a charged state
$|1,m-\frac{1}{2}\rangle$, i.e.,
\begin{eqnarray}
\mu^{\downarrow}_{\mathrm{S}}<
E_{|1,m-\frac{1}{2}\rangle}-E_{|0,m\rangle}.
\end{eqnarray}
Because of the anisotropy of SMM, the spectrum of the transitions
for all the possible $m$ has a finite width, as shown in Fig.
\ref{fig:transitionenergies}. Therefore, the splitting of
$\mu_{\uparrow}$ and $\mu_{\downarrow}$, i.e., the spin bias, must
be large enough to overcome this spectrum width. This thereby
defines a threshold spin bias for the reversal.

In the process of the reversal, only $\uparrow $ electrons enter SMM
while only $\downarrow $ electrons leave SMM at almost the identical
rate. As a result, a nearly pure spin current, instead of electric
current, flows between the lead and the SMM. Once the reversal is
accomplished, i.e., $|1,21/2\rangle $ is occupied, no more $\uparrow
$ ($\downarrow $) electron can enter (leave), and the pure spin
current decays to zero. The magnetization reversal mechanism
discussed above is supported by the simulation results shown in Fig.
\ref{fig:avalanche}.

Note that the neutral and charged branches employed in the above
reversal mechanism universally exist not only within the giant spin
approximation,\cite{Heersche2006prl96206801,Jo2006nl62014,Romeike2006prl96196805,Timm2006prb73235304,Elste2006prb73235305,Misiorny2007prb76054448}
but also in the multi-ion model\cite{Lehmann2007prl98117203} and
density functional
theory.\cite{Park2004.prb.70.054414,Michalak.arxiv.0812.1058} In the
multi-ion model of Lehmann and Loss,\cite{Lehmann2007prl98117203}
the case of ferromagnetic inter-ion interaction corresponds to the
type of SMM discussed in the present work. The green dash-dotted
lines in Fig. 1(a) of their paper describe the transitions between
the neutral and charged branches with the difference of the total
angular momentum by $1/2$. Recent density-functional theory
calculation also concludes that the total angular momenta for the
ground-state neutral and anionic branches are $10$ and $21/2$,
respectively.\cite{Michalak.arxiv.0812.1058} Therefore the above
reversal mechanism should be universally described by most models
proposed by far.

Later we will also show that the above mechanism is not
qualitatively affected by the transverse anisotropy (Sec.
\ref{sec:transverse}) and non-collinearity (Sec. \ref{sec:theta}).

\section{\label{sec:model}General formalism }

\subsection{Model}\label{sec:modelmodel}
In this work, we want to focus on the possibility of using spin
bias, and what we need are one neutral branch and one charged branch
as discussed above. Specifically, we adopt the model proposed by
Timm and Elste\cite{Timm2006prb73235304,Elste2006prb73235305} to
describe SMM, which consists of the lowest unoccupied molecular
orbital (LUMO) and the phenomenological giant spin (GS) $\mathbf{S}$
of the molecule,
\begin{eqnarray}\label{SMM}
H_{\mathrm{SMM}} &=&(\epsilon_0-eV_g)\sum_{\sigma \in \{\uparrow
,\downarrow
\}}n_{\sigma }+Un_{\uparrow }n_{\downarrow }\nonumber\\
&&-(D+\delta D\sum_{\sigma }n_{\sigma })S_{z}^{2}+H'
+E_{\mathrm{ion}}(V_g)\nonumber\\
&&-J\mathbf{s}\cdot \mathbf{S},
\end{eqnarray}
where the first two terms depict the LUMO, $n_{\sigma }=d_{\sigma
}^{\dag }d_{\sigma }$, and $d_{\sigma }$($d_{\sigma }^{\dag }$) are
the annihilation (creation) operators for the LUMO, whose on-site
energy is tunable by a gate voltage $V_g$. $-e$ is the electron
charge. $U$ is the on-site Coulomb repulsion.

The third and forth terms are for the anisotropy of GS, where $D$
describes the easy axis anisotropy and $\delta D$ accounts for the
correction to the easy axis anisotropy by the occupation of LUMO.
$H'$ describes the transverse anisotropy. We formally consider its
possible leading terms
\begin{eqnarray}
H' = B_2(S_+^2+S_-^2)+B_4(S_+^4+S_-^4).\nonumber\\
\end{eqnarray}
For Mn$_{12}$ac, there is usually only $B_4$ term. We also include
$B_2$ term because we want to investigate the effect of the
transverse terms in a general way. For Mn$_{12}$ac, $B_4$ are
several orders smaller than the easy axis anisotropy $D$ (Table 4.1
of Ref.\onlinecite{Gatteschi2006book}). Therefore, we assume that
the extra electron brings no correction to them. Following most
experiments and theories by
far,\cite{Heersche2006prl96206801,Jo2006nl62014,Romeike2006prl96196805,Timm2006prb73235304,Elste2006prb73235305,Lehmann2007prl98117203}
we assume the extra electrons change only the magnitude of the easy
axis anisotropy.

The fifth term is the energy of ions that form the giant spin, which
is also proportional to $V_g$ in the same way as LUMO. The last term
describes the Hund's rule coupling $J$ between the giant spin
$\mathbf{S}$ and the electron spin $\mathbf{s}$ in the LUMO.
$\mathbf{s}=\frac{1}{2}\sum d_{\alpha }^{\dag
}(\overrightarrow{\sigma})_{\alpha \beta }d_{\beta }$, where
$\overrightarrow{\sigma} $ are the vector of Pauli matrices.

The Hamiltonian for the electrodes used to probe SMM reads
\begin{equation}
H_{\mathrm{lead}}=\sum_{k,\alpha ,\tau }\epsilon_{k\alpha}c_{k\alpha
\tau }^{\dag }c_{k\alpha \tau },
\end{equation}
where $c_{k\alpha \tau }^{\dag }(c_{k\alpha \tau })$ is the creation
(annihilation) operator for a continuous state in the $\alpha
$($\in\{\mathrm{S,D}\}$) lead with energy $\epsilon_{k\alpha}$ and
spin $\tau\in\{+,-\} $. In reality, there should be an angle
$\theta$ between the easy axis of SMM and the spin orientation in
the electrodes, so the spin quantization in SMM is denoted
$\sigma\in \{\uparrow,\downarrow\}$, and in the lead as
$\tau\in\{+,-\}$. The operator for spin-$\tau$ electron in the lead
is related to spin-$\sigma$ operator through an SU(2) rotation.

The tunneling between the LUMO and the electrodes is described by
\begin{eqnarray}\label{HT}
H_{\mathrm{T}}&=&\sum_{k,\alpha }V_{k\alpha
}[(\cos\frac{\theta}{2}c_{k\alpha + }^{\dag
}-\sin\frac{\theta}{2}c_{k\alpha - }^{\dag })d_{\uparrow
}\nonumber\\
&&\ \ \ \ \ \ \ \ +(\sin\frac{\theta}{2}c_{k\alpha + }^{\dag
}+\cos\frac{\theta}{2}c_{k\alpha - }^{\dag
})d_{\downarrow }]+H.c.,\nonumber\\
\end{eqnarray}
where $\theta\in[0,\pi/2]$. We set the easy axis of the SMM as $z$
axis. For $\theta\in [\pi/2,\pi]$ one just reverses the positive
direction of $z$ axis.  In short, the total Hamiltonian for the
system we are studying is
\begin{eqnarray}
H_{\mathrm{total}}=H_{\mathrm{SMM}}+H_{\mathrm{lead}}+H_{\mathrm{T}}.
\end{eqnarray}

We believe that the Hamiltonian employed in this work captures the
physics required for the magnetization reversal, though it is a
simplified model. The physical picture of this model can be
understood as follows. In a $\mathrm{Mn}_{12}$ac molecule, eight
spin-2 $\mathrm{Mn}^{3+}$ ions and four spin-$\frac{3}{2}$
$\mathrm{Mn}^{4+}$ ions are exchange-coupled. As a good quantum
number, their total angular momentum may adopt many values, referred
as different branches. The branch with the lowest energy consists of
21 states with a total angular momentum $S=10$. The branches with
other values of total angular momentum are much higher in energy.
Due to the anisotropy along the easy axis, two degenerate ground
states of $S=10$ branch are the states with $z$ component of the
total angular momentum $S_z=\pm S$, respectively. In this sense the
system is regarded as a giant spin of $S=10$, and is simply
described by the term $-DS_z^2$. By adding an extra electron, the
energy of the molecule changes in several aspects: (i) The first is
the on-site and charging energy to add this electron. This is
described by $\epsilon_0\sum_{\sigma}n_{\sigma}$. If we set the
Fermi level of the lead as the reference point, this energy can be
compensated by the gate voltage, so it is absorbed into the term
$(\epsilon_0-eV_g)\sum_{\sigma}n_{\sigma}$. The $U$ term is due to
adding the second excess electron to the same LUMO. To simplify the
problem, we assume the second electron will also occupy the same
LUMO, and exclude the possibility to occupy other states. This is
purely theoretical simplification, and is believed not to affect
qualitatively the physical consequences in the present problem. We
have to emphasize that the energy of the ions
$E_{\mathrm{ion}}(V_g)$ is also tunable to the gate voltage $V_g$ in
the same way as the excess electron. It does not have to explicitly
appear in the Hamiltonian because it is counted in the energy of
each many-body state of the SMM, thus can be discarded. (ii) Second,
the added electron will interact with the giant spin to form spin
$S+1/2$ and $S-1/2$ branches of states. The energy difference
between these two branches can be characterized by the term
$-J\mathbf{S}\cdot \mathbf{s}$, where $J$ can be found by the
splitting between the two branches by using first-principle
calculation,\cite{Park2004.prb.70.054414} because of the splitting
is around $2JS$. (iii) The third is the correction of the anisotropy
due to the excess electron. Because the angular momentum of electron
spin is much smaller than that of the giant spin, the variation of
anisotropy in the presence of the excess electron, which mainly
leads to a curvature change in the energy spectrum, can be roughly
absorbed into the correction parameter $\delta D$.

Either the giant spin model proposed by Timm and
Elste\cite{Timm2006prb73235304,Elste2006prb73235305} or by Romeike
and co-workers\cite{Heersche2006prl96206801,Romeike2006prl96196805}
describes the above physical picture. Both characterize the
many-body eigen states by the electron occupation, the total angular
momentum (note that $\nu=\pm$ correspond to the total angular
momentum $=S\mp \frac{1}{2}$), and the $z$ component of total
angular momentum. When calculating the matrix elements $\langle i
|d_{\sigma}|j\rangle $, the Clebsh-Gordan
coefficients\cite{Romeike2006prl96196805} correspond to the linear
combination coefficients\cite{Timm2006prb73235304} $\alpha_m^{\pm},
\beta_m^{\pm}$. Most importantly, both model are able to capture the
main features of the experiments, e.g., the sophisticated magnetic
excitations and the negative differential conductance observed in
the experiments.\cite{Heersche2006prl96206801}

Besides, Lehmann and Loss\cite{Lehmann2007prl98117203} think that
the inclusion of the excess electron with respect to the uncharged
SMM should start with a multi-ion model, in which $N$ ion sites are
considered, each with an ion spin $s$. Nearest ions are coupled, by
either ferromagnetic or anti-ferromagnetic exchange interaction. The
excess electron can occupy and hop among any of these ion sites. The
Hund's rule coupling between the excess electron and each ion is
local, as well as the anisotropy. When considering
anti-ferromagnetic inter-ion coupling, the ground state adopts a
zero total angular momentum and apparently is not the case for
$\mathrm{Mn}_{12}$ac (but valid for other SMMs, such as
$\mathrm{Mn}_4$
dimer\cite{Wang1996Inorg.Chem35.7578,Hu2003.prb.68.104407}). The
spatial selection rules they predicted mainly occur for the
anti-ferromagnetic case, thus will not be considered in this work.
When considering ferromagnetic inter-ion coupling, the ground state
adopts a maximal angular momentum $Ns$. They considered only one
excess electron. In this case, the electron is free to hop and
couple to all the ions. As a result, the local Hund's rule coupling
and anisotropy give rise to global giant spin properties. In a word,
the single excess electron and ferromagnetic inter-ion case of
Lehmann and Loss's model shares the same spirit of those by Timm and
Elste and Romeike \emph{et al}.

The spatial selection rules are also predicted by using the density
functional theory.\cite{Michalak.arxiv.0812.1058} However, because
of the lead ($\sim 100$ nm) used in the
experiment\cite{Heersche2006prl96206801} is much wider than the size
of the molecule ($< 5$ nm), we think the spatial selection rules,
which need precise contacts between the lead and the ion sites of
the molecule, could be smeared in realistic samples.

Though the coexistence of electron-phonon interaction and magnetic
excitation is observed,\cite{Heersche2006prl96206801,Jo2006nl62014}
the phonon frequency is beyond the energy scale of the current work.
For example, the phonon frequency observed by Heersche \emph{et
al}.\cite{Heersche2006prl96206801} is about $14$ meV, while the
magnetic excitations
observed\cite{Heersche2006prl96206801,Jo2006nl62014} or in this work
(Fig. \ref{fig:transitionenergies}) and the spin bias (Fig.
\ref{fig:hysteresis}) are of order of meV. Therefore, we do not
consider electron-phonon interaction and its related effect in this
work, e.g., Franck-Cordon blockade\cite{Koch2005} or
thermal-activated effect induced by spin-phonon
interaction.\cite{Politi1995.prl.75.537}

\subsection{\label{sec:parameters} SMM states in absence of transverse anisotropy and parameters
for simulations}

We will use the eigen states of $H_{\mathrm{SMM}}$ when $B_2=B_4=0$
as unperturbed states.\cite{Timm2006prb73235304} The transverse
anisotropy will be taken into account by perturbation for small
$B_2$ and $B_4$.

When $B_2=B_4=0$, $H_{\mathrm{SMM}}$ leads to four branches of
states for the isolated SMM denoted by $|n,m\rangle ^{\nu }$, where
$n$($=0,1,2$) is the LUMO occupation and $m$ is the quantum number
for ($S_z+s_z$), the $z$ component of total angular
moment.\cite{Timm2006prb73235304} Degeneracy index $\nu $ only
appears when $n=1$. The four branches are:

The empty branch
\begin{equation}
|0,m\rangle \equiv |0\rangle_{\mathrm{LUMO}} \otimes
|m\rangle_{\mathrm{GS}},
\end{equation}
where $m\in [-S,S]$.

The two singly-occupied branches
\begin{equation}
|1,m\rangle^{\pm}\equiv
\alpha^{\pm}_m|\downarrow\rangle_{\mathrm{LUMO}}\otimes|m+\frac{1}{2}\rangle_{\mathrm{GS}}+\beta^{\pm}_m|\uparrow\rangle_{\mathrm{LUMO}}\otimes|m-\frac{1}{2}\rangle_{\mathrm{GS}},
\end{equation}
where $m\in [-S-\frac{1}{2},S+\frac{1}{2}]$ for $\nu=-$, $m\in
[-S+\frac{1}{2},S-\frac{1}{2}]$ for $\nu=+$.

The doubly-occupied branch
\begin{equation}
|2,m\rangle \equiv
|\uparrow\downarrow\rangle_{\mathrm{LUMO}}\otimes|m\rangle_{\mathrm{GS}},
\end{equation}
where $m\in [-S,S]$.

One can refer to Fig. 2 of Ref. \onlinecite{Misiorny2007prb76054448}
to have a direct impression of these four branches. But different
from Ref. \onlinecite{Misiorny2007prb76054448}, in this work the
higher two branches are far above the lower two branches because of
large $J$ and $U$.

We adopt the parameters based on recent experiments and
first-principles calculations for Mn$_{12}$ac ($S=10$) as $D=0.056$
meV, $\delta D=-0.008$ meV,\cite{Heersche2006prl96206801} and
$J=3.92$ meV.\cite{Park2004.prb.70.054414,Timm2007prb76014421}
Because $\epsilon_0$ can be compensated by $V_g$, we set
$\epsilon_0=0$ for convenience. We choose $U=25$meV, which is
comparable to the width of Coulomb diamond in
experiments.\cite{Heersche2006prl96206801,Jo2006nl62014} For the
above parameters (large $J$ and $U$), the two highest branches
$|2,m\rangle $ and $|1,m\rangle ^{+}$ are neglected in the present
work because the branch $|1,m\rangle ^{+}$ is about $2SJ$ (about
several tens of meV) above the branch $|1,m\rangle ^{-}$, and the
branch $|2,m\rangle$ is even higher. In the following numerical
simulations, we consider only the branches $|0,m\rangle $ and
$|1,m\rangle^{-}$. By choosing suitable gate voltage $V_g$, these
two branches can be nearly degenerate with respect to the Fermi
levels of the leads.\cite{Romeike2006prl96196805} A typical
situation of $V_g=-20$ mV is shown in Fig. \ref{fig:transitions}.

\subsection{Perturbative correction to SMM states by transverse anisotropy}

For Mn$_{12}$ac, the transverse anisotropies $B_2$ and $B_4$ are
several orders smaller than the easy axis anisotropy $D$ (Table 4.1
of Ref. \onlinecite{Gatteschi2006book}). For $B_2 \ll D/S^{2}$ and
$B_4 \ll D/S^{4}$, they can be taken into account by the standard
perturbation calculation. Note that degenerate states such as
$|0,\pm 1\rangle $ and $|0,\pm 2\rangle $ are coupled by $H'$, so
one has to perform a degenerate perturbation calculation. We
consider the first-order correction to the states and the
second-order to their energies (please refer to Appendix
\ref{sec:perturbation} for details).

In the presence of weak $B_2$ and $B_4$, the eigen states can only
be approximately labeled by the quantum number $m$ of $(S_z+s_z)$,
and becomes a linear combination of all the states with the same
LUMO occupation,\cite{Romeike2006prl96196805} i.e.,
\begin{eqnarray}\label{state_perturbed}
|0,m\rangle_p &=& \sum_{m'=-S}^S C_{m',m}^{0}|0,m'\rangle,\nonumber\\
|1,m\rangle^{-}_p &=&\sum_{m'=-S-\frac{1}{2}}^{S+\frac{1}{2}}
C_{m',m}^{-}|1,m'\rangle^{-}\nonumber\\
&&+\sum_{m'=-S+\frac{1}{2}}^{S-\frac{1}{2}}
C_{m',m}^{+}|1,m'\rangle^{+}
\end{eqnarray}
where $p$ stands for \emph{perturbed} states by $B_2$ and $B_4$. One
can expect that $C_{m,m}^{0,-}\sim 1$, i.e., $|0,m\rangle_p$ are
mainly contributed by $|0,m\rangle$, and $|1,m\rangle_p^-$ by
$|1,m\rangle^-$.

The projection of magnetization along $z$ axis for the perturbed
states are obtained, for the branch $|0,m\rangle_p$,
\begin{eqnarray}\label{m_perturb0}
m_p = \sum_{m'=-S}^S|C^{0}_{m',m}|^2m';
\end{eqnarray}
and for the branch $|1,m\rangle_p^-$,
\begin{eqnarray}\label{m_perturb1}
m_p = \sum_{m'=-S-\frac{1}{2}}^{S+\frac{1}{2}}|C^{-}_{m',m}|^2m'+
\sum_{m'=-S+\frac{1}{2}}^{S-\frac{1}{2}}|C^{+}_{m',m}|^2m',
\end{eqnarray}
and $m_p=m$ when $B_2=B_4=0$.

According to Eqs. (\ref{psi1}) and (\ref{E2}), the validity of the
perturbation requires that
\begin{eqnarray}
\frac{H'_{ji}}{E_i-E_j} &\ll & 1,
\end{eqnarray}
for arbitrary $i,j\in\{|0,m\rangle,|1,m\rangle^-\}$. By using Eqs.
(\ref{E00}) and (\ref{H0p}), one can estimate that $H'_{ji}$ can be
as large as $B_2S^2$ and $B_4 S^4$ when $m\sim 0$, and $E_i-E_j$ can
be as small as $D$ when $m\sim 0$. Therefore, the perturbation only
applies for $B_2\ll D/S^2$ and $B_4 \ll D/S^4$, which are also
reasonable values for realistic Mn$_{12}$ac molecules (Table 4.1 of
Ref. \onlinecite{Gatteschi2006book}).

\subsection{Pauli rate equations}

When connected to the leads, the eigen states of $H_{\mathrm{SMM}}$
can transit to each other by exchanging electrons with the lead. In
the weak-coupling regime and when neutral and charged states are
nearly degenerate, the sequential tunneling is dominant. The
transitions are well described by the Pauli rate equations of a
reduced density matrix spanned by the eigen states of SMM. In this
approach, Born approximation and Markoff approximation are employed,
and $H_{\mathrm{T}}$ is treated by perturbation up to the second
order.\cite{Blum1996book} (Please refer to Appendix
\ref{sec:rateappendix} for details.) The rate equation can be
expressed in a compact form,
\begin{eqnarray}\label{rateequation}
\partial _{t}P_{i}=\sum_{j}R_{ij}P_{j},
\end{eqnarray}
where $0 \leq P_{i} \leq 1$ are the probability to find the state
$i$. In this work, $i$ or $j$ belongs to the $2S+1=21$ states from
the branch $|0,m\rangle_p$ and $2(S+\frac{1}{2})+1=22$ states from
the branch $|1,m\rangle^-_p$, the off-diagonal and diagonal terms of
the coefficient matrix are given by
\begin{eqnarray}
R_{i\neq j}=\sum_{\alpha \sigma }R_{j\rightarrow i}^{\alpha \sigma
},\ \ \ \ R_{ii}=-\sum_{j\neq i}\sum_{\alpha \sigma }R_{i\rightarrow
j}^{\alpha \sigma }
\end{eqnarray}
where
\begin{eqnarray}\label{ratematrix}
R_{i\rightarrow j}^{\alpha \uparrow } &=&\Gamma\{|\langle
i|d_{\uparrow }|j\rangle
|^{2}[\cos^2(\theta/2)f(E_{j}-E_{i}-\mu_{\alpha }^{+
})\nonumber\\
&&\ \ \ \ \ \ \ \ \ \ \ \ \ \ \ \
+\sin^2(\theta/2)f(E_{j}-E_{i}-\mu_{\alpha }^{-})]
\nonumber \\
&&\ +|\langle j|d_{\uparrow }|i\rangle
|^{2}[\cos^2(\theta/2)f(E_{i}-E_{j}+\mu _{\alpha
}^{+})\nonumber\\
&&\ \ \ \ \ \ \ \ \ \ \ \ \ \ \ \
+\sin^2(\theta/2)f(E_{i}-E_{j}+\mu _{\alpha }^{-})]\}.\nonumber\\
\end{eqnarray}
where the Fermi distribution $f(x)=1/[\exp (x/k_{B}T)+1]$ is
spin-resolved. The coupling between LUMO and the $\alpha $ lead is
assumed to be a constant parameter $\Gamma _{\alpha }^{\tau }=2\pi
\sum_{k}|V_{k\alpha }|^{2}\delta (\omega -\epsilon _{k\alpha
})=\Gamma $ for nonmagnetic lead. One just replaces $\uparrow$ by
$\downarrow$ and exchanges $+$ and $-$ to obtain $R_{i\rightarrow
j}^{\alpha \downarrow }$. Because the drain lead is non-magnetic and
not subjected to the spin bias, one can assume
$\mu_{\mathrm{D}}^{+/-}=\mu_{\mathrm{D}}^{\uparrow/\downarrow}$ for
simplicity. For the source,
$\mu_{\mathrm{S}}^{+/-}=\mu_{\mathrm{S}}^{\uparrow/\downarrow}$ only
when $\theta=0$.

\begin{figure}[tbp]
\centering\includegraphics[width=0.45\textwidth]{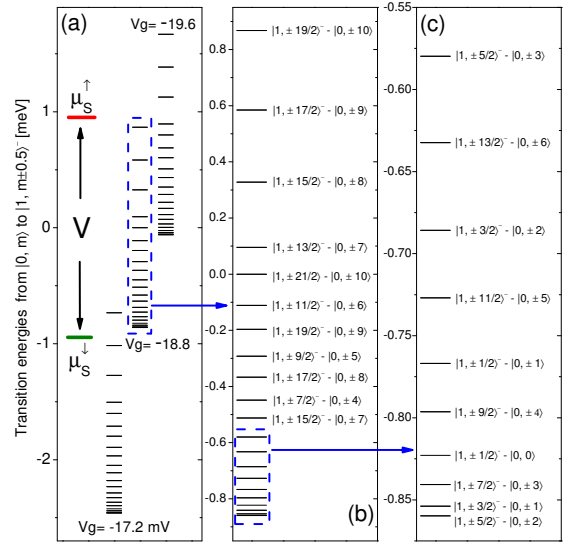}
\caption{(a) Resonant energies to add an extra electron into SMM via
transitions from state $|0,m\rangle $ to $|1,m\pm \frac{1}{2}\rangle
^{-}$ for three different $V_g$. The middle point of
$\mu_{S}^{\uparrow/\downarrow}$ is set as energy zero point so that
$\mu_{S}^{\uparrow/\downarrow}=\pm V/2$. (b) Zoom-in when
$V_g=-18.8$ mV, where the middle point of the entire spectrum is
aligned with 0, i.e. the middle point of the Fermi levels
$\mu_{S}^{\uparrow/\downarrow}$. (c) Zoom-in of the lowest
transitions in the middle panel. The notation $|1,m\pm
\frac{1}{2}\rangle ^{-}-|0,m\rangle $
is short for $E_{|1,m\pm \frac{1}{2}\rangle ^{-}}-E_{|0,m\rangle }$. Note that $%
E_{|1,-(m\pm \frac{1}{2})\rangle ^{-}}-E_{|0,-m\rangle }=E_{|1,m\pm \frac{1}{%
2}\rangle ^{-}}-E_{|0,m\rangle }$ are degenerate in the absence of
magnetic field. Basic parameters are given in Sec.
\ref{sec:parameters}.} \label{fig:transitionenergies}
\end{figure}

\begin{figure}[tbph]
\centering\includegraphics[width=0.45\textwidth]{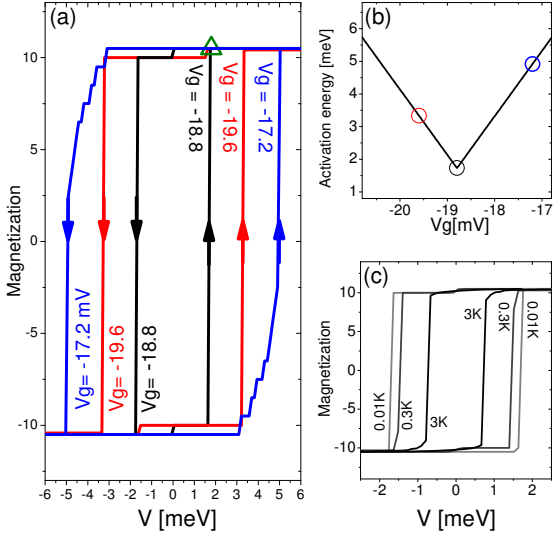}
\caption{(a) Magnetic hysteresis loops for different gate voltage
$V_g$, when scanning the spin bias $V$ back and forth. $T=0.01$K.
Arrows indicate the scanning direction of $V$. The scanning is
assumed to be slow enough to allow the system relax to steady state.
The triangle corresponds to the magnetic reversal point for the
simulation in Figs. \ref{fig:avalanche} and \ref{fig:avalanche
probabilities}. (b) The activation energy vs $V_g$. It consists of
two slopes connecting at $V_g=-18.8 $ mV. The left slope is
determined by the transition energy $E_{|1,\pm 19/2\rangle
^{-}}-E_{|0,\pm 10\rangle }$, the right slope is determined by the
transition energy $E_{|1,\pm 5/2\rangle ^{-}}-E_{|0,\pm 2\rangle }$.
Circles correspond to $V_g$=-17.2, -18.8, and -19.6 mV in the left
panel. (c) Temperature-dependent magnetic hysteresis loops as a
function of $V$ for $V_g=-18.8$ mV. Results for experimental
temperatures 3K (Ref. \onlinecite{Heersche2006prl96206801}) and 0.3K
(Ref. \onlinecite{Jo2006nl62014}) are presented. Other parameters
are given in Sec. \ref{sec:parameters}.} \label{fig:hysteresis}
\end{figure}

All the physical quantities can be expressed in terms of $P_{i}$,
such as LUMO occupation $\sum_{i}n_{i}P_{i}$, the SMM magnetization
$\sum_{i}m_p^iP_{i}$, and the $\sigma $ current flowing from the
$\alpha $ lead to SMM
\begin{eqnarray}
I_{\alpha }^{\sigma
}=-e\sum_{ij}(n_{i}-n_{j})R^{\alpha\sigma}_{j\rightarrow i}P_{j},
\end{eqnarray}
where $n_{i}$ and $m_p^i$ correspond to the $n$ and $m_p$ in the
state $i\equiv |n,m\rangle^{\nu}_p $.

Notice that since we consider only the branches $|0,m\rangle$ and
$|1,m\rangle^- $, the terms such as $\langle i | d_{\sigma}|k\rangle
\langle j|d_{\sigma}^{\dag}|i\rangle$ in the general formalism for
sequential tunneling\cite{Lehmann2007prl98117203} lead to $k=j$,
which remove the off-diagonal terms $P_{i\neq j}$ from the equation
of motion for $P_{i}$ and from the current formula, and lead to the
Pauli type rate equations of only diagonal terms. (Please refer to
Appendix \ref{sec:rateappendix} for details.)

Although the above Pauli rate equations formalism is widely employed
for the molecules weakly coupled to the
electrodes,\cite{Romeike2006prl96196805,Timm2006prb73235304,Elste2006prb73235305,Timm2007prb76014421,Elste2007prb75195341,Misiorny2007epl7827003,Misiorny2007prb75134425,Misiorny2007prb76054448}
its validity in this work deserves some discussion. In the
weakly-coupled regime, i.e., the lead-molecule coupling $\Gamma$ is
so small that between two consecutive electron tunnelings ($\sim
2/\Gamma$), there is enough time for the molecule to relax to the
eigen states of $H_{\mathrm{SMM}}$. For example, suppose the
molecule is originally at the state $|0,m-\frac{1}{2}\rangle \equiv
|0\rangle_{\mathrm{LUMO}}\otimes|m-\frac{1}{2}\rangle_{\mathrm{GS}}$.
The molecule will transit to the state
$|\uparrow\rangle_{\mathrm{LUMO}}\otimes|m-\frac{1}{2}\rangle_{\mathrm{GS}}$
after a spin-up electron tunnels in. Note that
$|\uparrow\rangle_{\mathrm{LUMO}}\otimes|m-\frac{1}{2}\rangle_{\mathrm{GS}}$
is not an eigen state of $H_{\mathrm{SMM}}$, it will relax to the
eigen state $|1,m\rangle^{-}\equiv
\alpha^{-}_m|\downarrow\rangle_{\mathrm{LUMO}}\otimes|m+\frac{1}{2}\rangle_{\mathrm{GS}}+\beta^{-}_m|\uparrow\rangle_{\mathrm{LUMO}}\otimes|m-\frac{1}{2}\rangle_{\mathrm{GS}}$.
The time scale of this relaxation is $\sim 1/J$, since it is the
term $-J\mathbf{s}\cdot \mathbf{S}$ that couples the states
$|\downarrow\rangle_{\mathrm{LUMO}}\otimes|m+\frac{1}{2}\rangle_{\mathrm{GS}}$
and
$|\uparrow\rangle_{\mathrm{LUMO}}\otimes|m-\frac{1}{2}\rangle_{\mathrm{GS}}$.
To make sure that the relaxation happens before the next electron
tunneling event, $J\gg\Gamma $ must be satisfied. Therefore, in this
work we use $J=3.92$ meV from the first principles
calculation\cite{Park2004.prb.70.054414,Timm2007prb76014421} and
$\Gamma \sim 0.01$ meV.

\section{\label{sec:numericals}magnetization reversal and detection using spin bias}

In this section, we will present the numerical results when
$\theta=0$ and $B_2=B_4=0$. The cases when $B_2,B_4\ll D$ and
$\theta\neq0$ will be presented in Secs. \ref{sec:transverse} and
\ref{sec:theta}, respectively. We will show both cases bring no
qualitative change to the results presented in this section. When
$\theta=0$ and $B_2=B_4=0$, for convenience, the notations become
$\mu_{\mathrm{S}}^{\pm}\rightarrow
\mu_{\mathrm{S}}^{\uparrow/\downarrow}$ and
$|n,m\rangle^{\nu}_p\rightarrow |n,m\rangle^{\nu}$.

We will first introduce the magnetization reversal induced by spin
bias, including hysteresis loops as a function of spin bias in Sec.
\ref{sec:hysteresis}, and avalanche dynamics near the reversal point
in Sec. \ref{sec:avalanche}. Finally, we will discuss the
nondestructive measurement to the giant spin orientation using the
spin bias in Sec. \ref{sec:measurement}.

As discussed in Sec. \ref{sec:parameters}, our numerical simulations
are based on the Pauli rate equations for the 43 states of branches
$|0,m\rangle$ and $|1,m\rangle^-$. The parameters are already given
in Sec. \ref{sec:parameters}. We numerically solve Eq.
(\ref{rateequation}) using Runge-Kutta method with the relative
error smaller than $10^{-10}$. The steady solutions are obtained by
relaxing the equations until for each state, its relative
probability change with respect to the last iteration step is less
than $10^{-7}$.

We have checked the results for total 84 states. For the energy
scale of this work, the higher two branches $|1,m\rangle^+$ and
$|2,m\rangle$ are hardly occupied. Therefore, the abandonment of
these two branches is justified.

\subsection{\label{sec:hysteresis}Gate voltage tunable magnetic
hysteresis loops}

When $\theta=0$ and $B_2=B_4=0$, the rate equations expressed by Eq.
(\ref{rateequation}) indicate that, besides the selection rules
$|n-n^{\prime }|=1$ and $\ |m-m^{\prime }|=\frac{1}{2}$, to trigger
a transition $|0,m\rangle \rightarrow |1,m+\sigma \rangle ^{-}$ that
adds a $\sigma $ electron onto LUMO, two conditions are required:

(C1) Energetic requirement must be satisfied so that $\mu _{\alpha
}^{\sigma }>E_{|1,m+\sigma \rangle ^{-}}-E_{|0,m\rangle }$;

(C2) The state $|0,m\rangle $ should be occupied.

Similarly, to trigger a transition $|1,m+\sigma \rangle
^{-}\rightarrow |0,m\rangle $ that removes a $\sigma $ electron from
LUMO, it is required that:

(C3) Energetic requirement must be satisfied so that $\mu _{\alpha
}^{\sigma }<E_{|1,m+\sigma \rangle ^{-}}-E_{|0,m\rangle }$;

(C4) The state $|1,m+\sigma \rangle $ should be occupied.

To generate the charging-discharging sequence from $|1,-21/2\rangle
$ through $|1,21/2 \rangle $ shown in Fig. \ref{fig:transitions},
the energetic requirements (C1) must be satisfied for each of them.
We list all the 42 transition energies $E_{|1,m\pm
\frac{1}{2}\rangle ^{-}}-E_{|0,m\rangle }$ in Fig.
\ref{fig:transitionenergies}. Note that for the present model and
parameters the low and high bounds of the entire spectrum happen to
be $E_{|1,\pm 19/2 \rangle ^{-}}-E_{|0,\pm 10\rangle }$ and
$E_{|1,\pm 5/2\rangle ^{-}}-E_{|0,\pm 2\rangle }$, respectively.
Note that the highest and the lowest transition energies may differ
from sample to sample, but the following qualitative results are
unaffected. The reversal $S_z=-10 \rightarrow 10 $ requires $V$
large enough to satisfy
\begin{eqnarray}
\mu _{\mathrm{S}}^{\uparrow }&=&\frac{V}{2}>E_{|1,\pm 19/2\rangle
^{-}}-E_{|0,\pm 10\rangle }\nonumber\\
\mu _{\mathrm{S}}^{\downarrow }&=&-\frac{V}{2}<E_{|1,\pm 5/2\rangle
^{-}}-E_{|0,\pm 2\rangle },
\end{eqnarray}
which thereby defines a threshold voltage\cite{Timm2006prb73235304}
or activation energy.\cite{Misiorny2007prb76054448} Fig.
\ref{fig:transitionenergies} indicates clearly that the activation
energy is determined not only by the highest $E_{|1,\pm 19/2\rangle
^{-}}-E_{|0,\pm 10\rangle
}$,\cite{Timm2006prb73235304,Misiorny2007prb76054448} but also by
the lowest $E_{|1,\pm 5/2\rangle ^{-}}-E_{|0,\pm 2\rangle }$. More
importantly, as shown in Fig. \ref{fig:transitionenergies} the
entire spectra of all 42 transition energies can be shifted with
respect to $\mu_{\mathrm{S}}^{\uparrow /\downarrow }$ by tuning
$V_g$, which means that the activation energy is tunable by the gate
voltage. The activation energy as a function of $V_g$ is shown in
Fig. \ref{fig:hysteresis}(b). The activation energy can be minimized
when the center of the entire spectrum is aligned with 0, which is
about $V_g=-18.8$ mV for the present parameters. When $V_g<-18.8$,
e.g., $V_g=-19.6$, the activation energy is determined by $E_{|1,\pm
19/2\rangle ^{-}}-E_{|0,\pm 10\rangle }$ . When $V_g>-18.8$, e.g.,
$V_g=-17.2$, the activation energy is determined by $E_{|1,\pm
5/2\rangle ^{-}}-E_{|0,\pm 2\rangle }$. Using similar analysis, one
knows that the reversal $S_z=10 \rightarrow -10 $ requires
\begin{eqnarray}
\mu _{\mathrm{S}}^{\downarrow }&=&-\frac{V}{2}>E_{|1,\pm 19/2\rangle
^{-}}-E_{|0,\pm 10\rangle }\nonumber\\
\mu _{\mathrm{S}}^{\uparrow }&=&\frac{V}{2}<E_{|1,\pm 5/2\rangle
^{-}}-E_{|0,\pm 2\rangle },
\end{eqnarray}
As a result, the magnetization when sweeping $V$ back and forth must
exhibit a hysteresis loop. As shown in Fig. \ref{fig:hysteresis}(a),
the hysteresis loop varies with $V_g$ because the activation energy
is tunable by the gate voltage. When considering the broadening of
the Fermi surface at higher temperatures, thermal fluctuation will
activate the magnetic reversal before $V$ reaches exactly the
required activation energy. As a consequence, the magnetic
hysteresis loop shrinks when the temperature increases [Fig.
\ref{fig:hysteresis}(c)].

Note that the existence of the activation energy and its tunability
to the gate voltage only results from that the spectrum of all the
transition energies has a finite width, and the spectrum must has a
finite width because of the anisotropy. Therefore, although in the
reality the highest and lowest transition energies may differ from
sample to sample, the above results are qualitatively unaffected.

\subsection{\label{sec:avalanche}Avalanche dynamics at magnetic
reversal point}

\begin{figure}[tbph]
\centering\includegraphics[width=0.5\textwidth]{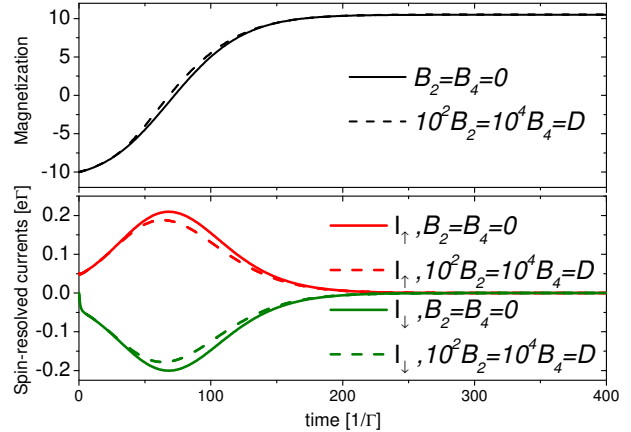}
\caption{The time-dependent magnetization and the spin-resolved
currents during the magnetic reversal. Solid and dashed lines
represent the cases in the absence and the presence of a weak
transverse anisotropy, respectively. The initial state is $
P_{|0,-10\rangle }=1$ and $P_{i\neq |0,-10\rangle }=0$. $T=0.01$K,
$V_g=-18.8$ mV, $V=1.8$ meV [the triangle in Fig.
\ref{fig:hysteresis}(a)]. Other parameters are given in Sec.
\ref{sec:parameters}. The positive sign of current stands for
flowing from source to SMM, and negative for from SMM to source. }
\label{fig:avalanche}
\end{figure}
\begin{figure}[tbph]
\centering\includegraphics[width=0.45\textwidth]{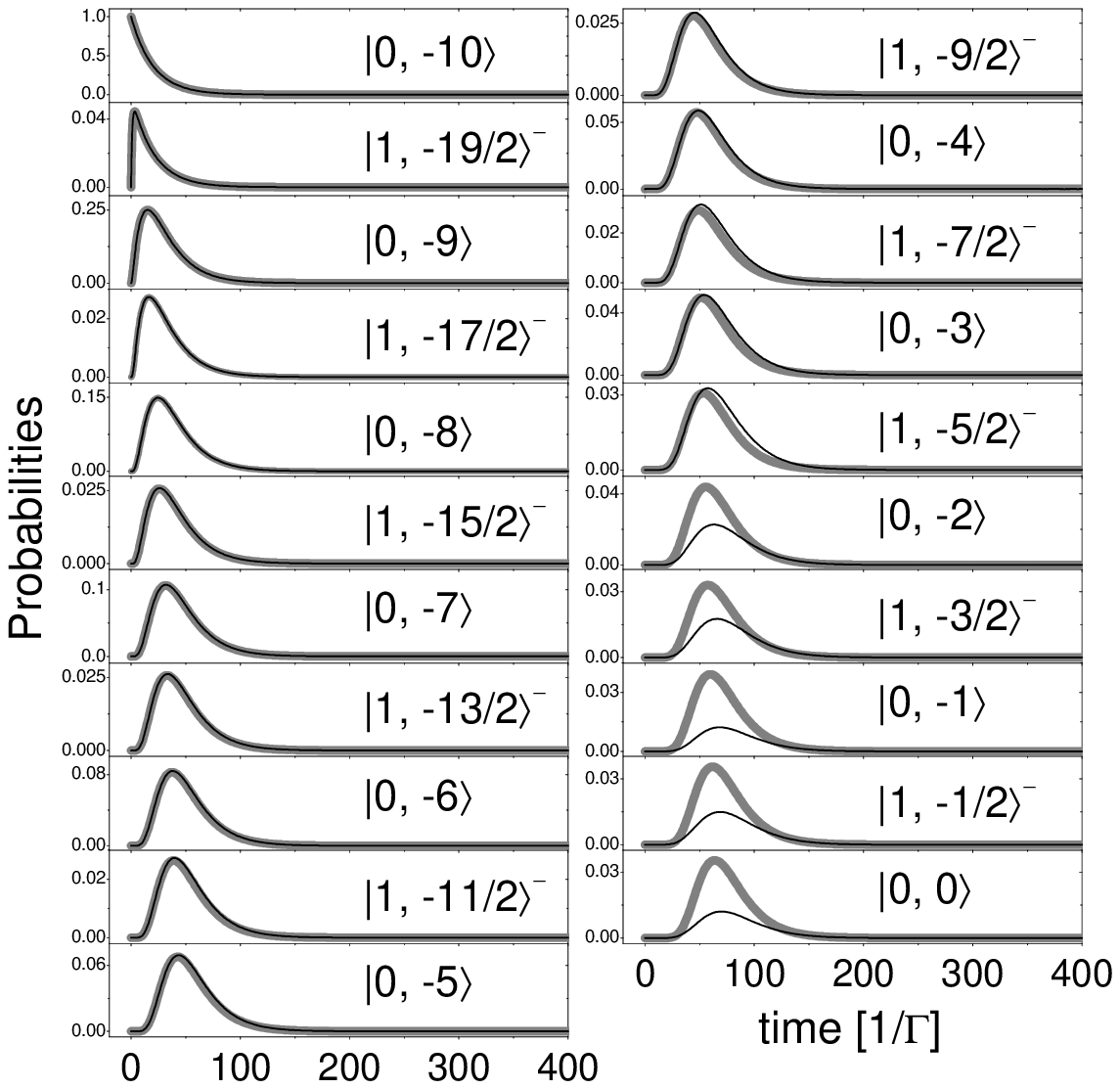}
\centering\includegraphics[width=0.45\textwidth]{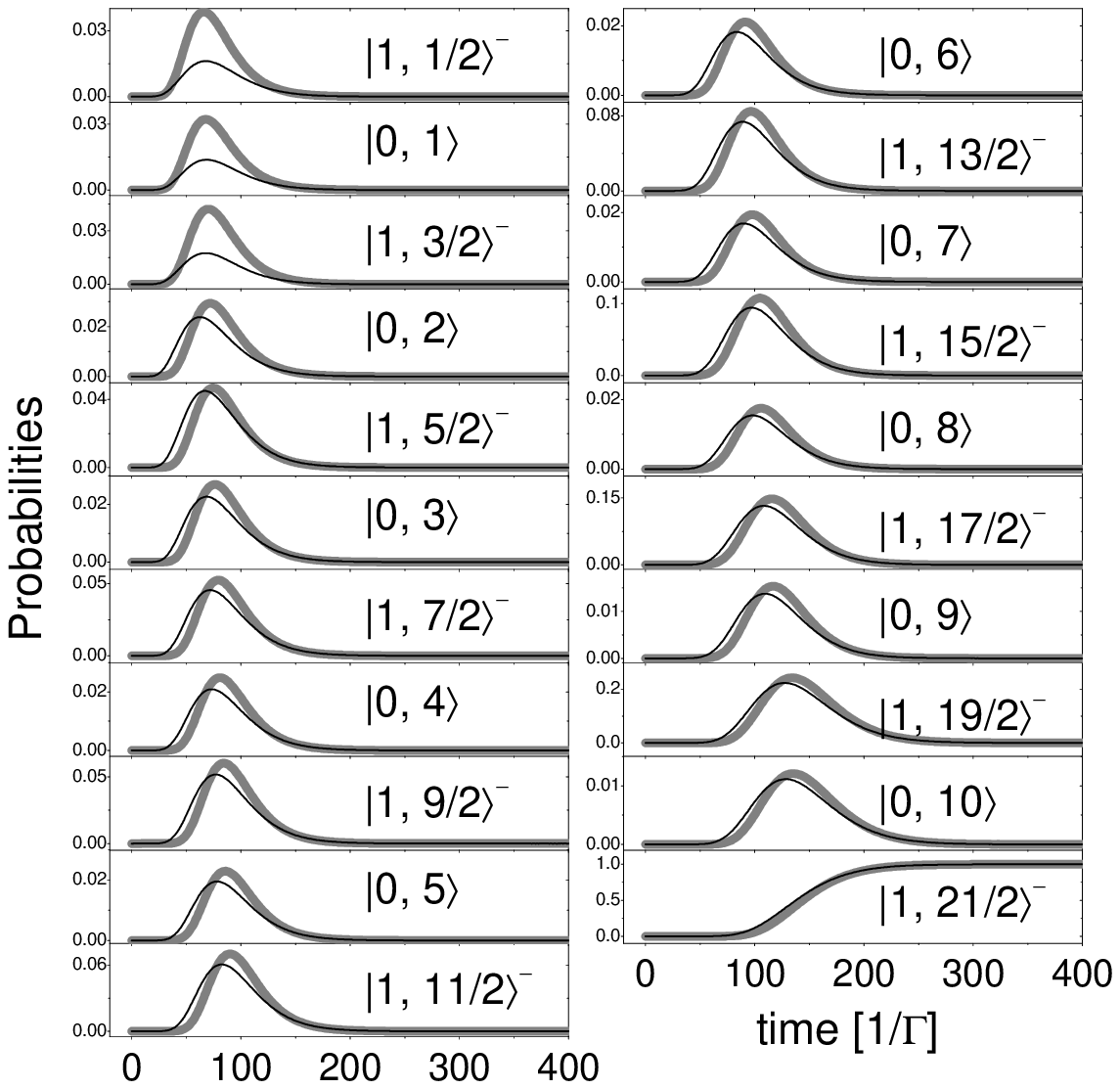}
\caption{During the magnetic reversal, the time-dependent
probabilities of the branches $|0,m\rangle$ and $|1,m\rangle^-$.
Thick and thin lines represent the cases when $B_2=B_4=0$ and
$10^2B_2=10^4B_4=D$, respectively. Note that the states from
$|0,-2\rangle$ through $|0,2\rangle$ are explicitly reshaped by the
$B_2$ and $B_4$. All the parameters are the same as Fig.
\ref{fig:avalanche}. } \label{fig:avalanche probabilities}
\end{figure}

We investigate the dynamics at the reversal point marked by the
triangle in Fig. \ref{fig:hysteresis}(a). On the left side of this
point, the activation energy is determined by $E_{|1,\pm 19/2
\rangle ^{-}}-E_{|0,\pm 10\rangle }$, and $V$ is smaller than
$2(E_{|1,-19/2\rangle ^{-}}-E_{|0,-10\rangle })$ while larger than
the rest 40 transition energies. Therefore, the reversal is blocked
at the state $|0,-10\rangle $ only because $|1,-19/2\rangle ^{-}$
can not be occupied. Once $|1,-19/2\rangle ^{-}$ is occupied when
$V$ exceeds $2(E_{|1,-19/2\rangle ^{-}}-E_{|0,-10\rangle })$, an
avalanche of the rest 40 transitions will be triggered. In Figs.
\ref{fig:avalanche} and \ref{fig:avalanche probabilities}, we
demonstrate the numerical simulation of this avalanche by showing
the time-dependent probabilities for the states of the branches
$|0,m\rangle$ and $|1,m\rangle^-$ when $V_g=-18.8$ mV and $V=1.8$
meV $>2(E_{|1,-19/2\rangle ^{-}}-E_{|0,-10\rangle })$. As we see,
all the intermediate states between $|0,-10\rangle $ and
$|1,21/2\rangle $ show a clear time sequence. Each of them is first
occupied, then reaches a maximum, and finally decays to zero. The
time scale of the avalanche process can be estimated by $\sim
300\times \frac{\hbar }{\Gamma }$, only $\sim$ 6 ns if we choose the
experiment fitting parameter\cite{Jo2006nl62014} $\Gamma
=8\mathrm{GHz}\approx0.033$ meV.

\subsection{\label{sec:measurement}Nondestructive detection to giant
spin orientation}

\begin{figure}[tbp]
\centering\includegraphics[width=0.5\textwidth]{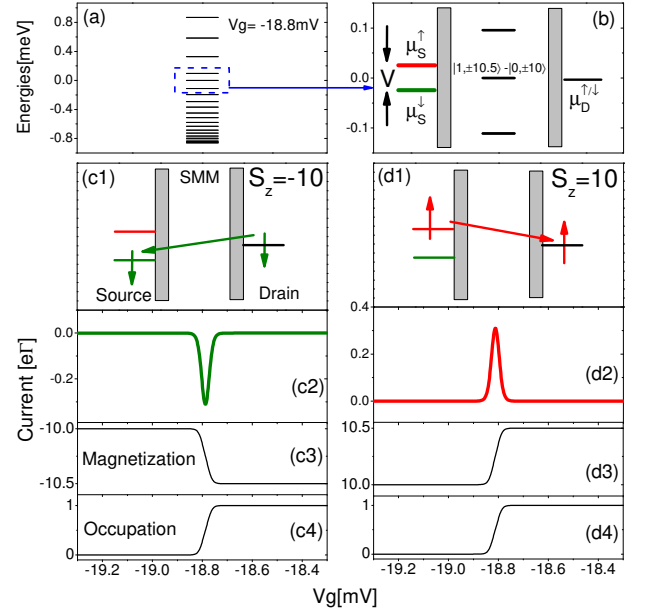}
\caption{Nondestructive detection to the giant spin orientation. (a)
Transition energies $E_{|1,m\pm \frac{1}{2}\rangle
^{-}}-E_{|0,m\rangle }$ when $V_g=-18.8$mV. (b) Zoom-in of (a) near
$E_{|1,\pm 21/2\rangle} -E_{|0,\pm 10\rangle }$. By tuning the gate
voltage $V_g$, the Fermi levels of leads
$\mu_{\mathrm{S/D}}^{\uparrow/\downarrow}$ can be located aligned
with the two degenerate transition energies $E_{|1,\pm 21/2\rangle
^{-}}-E_{|0,\pm 10\rangle }$. [(c1) and (d1)]: Schematics of how
fully polarized electric current flows when $S_{z}=\pm 10$.
[(c2)-(c4) and (d2)-(d4)]: For $S_{z}=\pm 10$, numerical results of
source-to-drain current, SMM magnetization, and LUMO occupation as
functions of $V_g$ in the presence of a small $V=0.05$meV. $T$=0.1K.
Other parameters are given in Sec. \ref{sec:parameters}. }
\label{fig:measure}
\end{figure}

A scheme to read out the SMM magnetization has been proposed by Timm
and Elste,\cite{Timm2006prb73235304} in which however the readout is
accompanied by the decay of SMM magnetization. Here we propose a
detection scheme, in which the giant spin of a SMM has already been
prepared to be at either $S$ or $-S$ ground state. Our goal is to
detect at which of the two orientations the giant spin points, and
most importantly, without destroying the giant spin orientation, by
means of spin bias.\cite{Lu2008prb77.235309} We have shown in Fig.
\ref{fig:avalanche} that, in the presence of only one lead only a
burst of pure spin current can flow only during the magnetic
reversal, i.e. one lead is not enough to maintain a steady current.
Thus, we introduce two leads for this detection, as shown in Fig.
\ref{fig:measure}(b). A small spin bias $V=0.05$ meV is applied to
only the source lead so that $\mu _{\mathrm{S}}^{\uparrow
/\downarrow }=\pm \frac{V}{2}$, while the Fermi level for the drain
lead $\mu _{\mathrm{D}}^{\downarrow /\uparrow }$ is set at 0. By
tuning the gate voltage $V_g$, the Fermi levels of leads
$\mu_{\mathrm{S/D}}^{\uparrow/\downarrow}$ can be located aligned
with the two degenerate transition energies $E_{|1,\pm
21/2\rangle^{-}}-E_{|0,\pm 10\rangle }$. Taking Figs.
\ref{fig:measure}(c1)-(c4), for example, the giant spin orientation
is initiated at $S_{z}=-10$, i.e., either $ |1,-21/2\rangle $ or
$|0,-10\rangle $ is likely to be occupied. On the left side of Figs.
\ref{fig:measure}(c2)-(c4), the molecule is at the state
$|0,-10\rangle$; on the right side of Figs.
\ref{fig:measure}(c2)-(c4), the molecule is at $|1,-21/2\rangle$.
Notice that the transition between them involves only $\downarrow $
electron. As shown in Fig. \ref{fig:measure}(b), when
\begin{eqnarray}
\mu_{\mathrm{S}}^{\downarrow}<E_{|1,\pm 21/2\rangle ^{-}}-E_{|0,\pm
10\rangle } \leq \mu_{\mathrm{D}}^{\downarrow},
\end{eqnarray}
a spin-down electron can be injected from the drain lead, inducing
the transition $ |0,-10\rangle \rightarrow |1,-21/2\rangle $, then
leaks to the source lead, making the molecule recover to the state $
|0,-10\rangle $. Such process repeats continuously, leading to a
steady fully polarized spin-down electric current flowing from the
drain to the source lead. Since we define the direction of current
from source to drain as positive, the spin-$\downarrow $ current is
negative, as shown by the negative peak in Fig.
\ref{fig:measure}(c2).

For the same situation shown in Fig. \ref{fig:measure}(b), if the
giant spin is oriented along $S_{z}=10$ as shown in Figs.
\ref{fig:measure}(d1)-(d4), only the transition between
$|1,21/2\rangle $ and $|0,10\rangle $ is possible, which will
generate a positive $\uparrow $ current from source to drain.
Because of the degeneracy of $E_{|1,21/2\rangle ^{-}}-E_{|0,
10\rangle }$ and $E_{|1,-21/2\rangle ^{-}}-E_{|0, -10\rangle }$, the
spin polarization and flowing direction of the steady current depend
only on the giant spin orientation.

Notice that the detection does not change the giant spin orientation
from initiated $S_{z}=\pm 10$ to other values. Specifically, in
order to destroy the state initiated $S_z=-10 $, the state $|0,
-10\rangle $ has to transit to $|1, -19/2\rangle ^{-}$ first, which
requires two conditions simultaneously: (1) $\mu
_{\mathrm{S/D}}^{\uparrow }>E_{|1,  -19/2\rangle ^{-}}-E_{|0,
-10\rangle }$. (2) $|0, -10\rangle $ is occupied. According to Fig.
\ref{fig:transitionenergies}, to make $\mu_{ \mathrm{S/D}}^{\uparrow
}>E_{|1, -19/2\rangle ^{-}}-E_{|0, -10\rangle }$, $V_g$ must be
$>$-18.8 mV. However, when $V_g>-18.8$, only $|1, -21/2\rangle $ is
allowed to be and fully occupied, instead of $|0, -10\rangle $, as
shown by the occupation and magnetization in Fig.
\ref{fig:measure}(c3) and (c4). Therefore, our detection scheme is
nondestructive.

\section{\label{sec:transverse}Effect of transverse anisotropy}

\begin{figure}[htbp]
\centering
\includegraphics[width=0.5\textwidth]{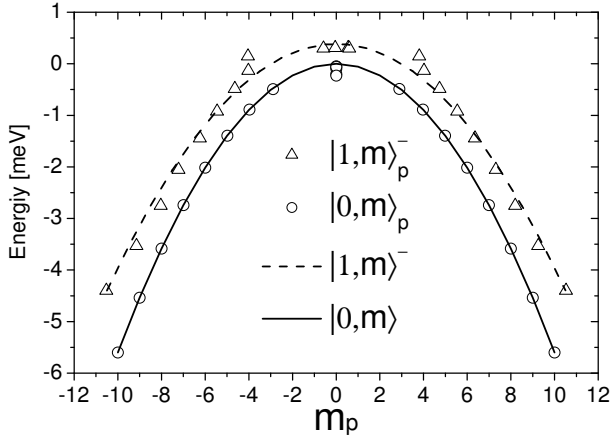}
\caption{The energies of the perturbed branches $|0,m\rangle_p$ and
$|1,m\rangle^-_p$ as a function of perturbed projection $m_p$.
Circle and triangle: $B_2=10^{-2}D$, $B_4=10^{-4}D$. Solid and
dashed lines: $B_2=B_4=0$. Other parameters are the same as Fig.
\ref{fig:transitions}. } \label{fig:band perturbed}
\end{figure}

\begin{figure}[tbph]
\centering\includegraphics[width=0.45\textwidth]{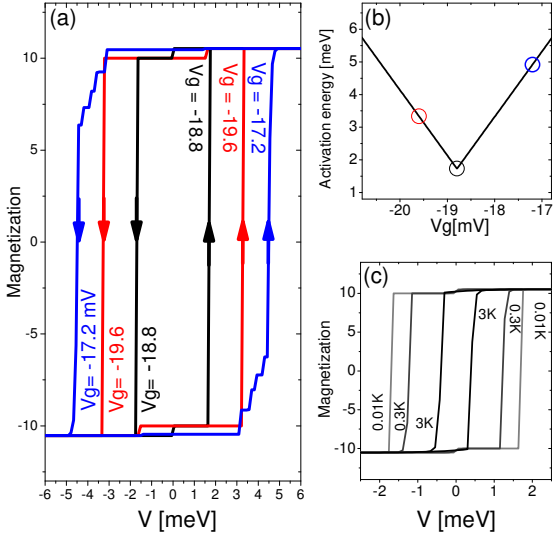}
\caption{The same as Fig. \ref{fig:hysteresis} except $B_2=10^{-2}D$
and $B_4=10^{-4}D$. }\label{fig:hysperturb}
\end{figure}

In this section, we will show that the small transverse anisotropy
only weakly affects the three main results presented in Sec.
\ref{sec:numericals}, and brings no qualitative change.

The transverse anisotropy leads to two main corrections of the SMM
eigen states. The first is the quantitative correction to the energy
and the projection of the magnetization along $z$ axis $m_p$
[defined by Eqs. (\ref{m_perturb0}) and (\ref{m_perturb1})]. The
second is the weak violation to the spin selection
rules\cite{Romeike2006prl96196805} according to Eq.
(\ref{state_perturbed}) so that transitions between states
$|m_p-m'_p|>1/2$ now are possible.

In Fig. \ref{fig:band perturbed}, we show the energy as a function
of the projection of magnetization along $z$ axis $m_p$ for each SMM
states, when $B_2=10^{-2}D$ and $B_4=10^{-4}D$. The case when
$B_2=B_4=0$ is also plotted for comparison. As we see, the changes
in energy and $m_p$ are ignorably small for those states with
$|m_p|\sim S$. However, the states with $|m_p|\sim 0$ are greatly
reshaped, and in some extent mixed together.

The influence of this mixture on the reversal dynamics is
demonstrated by the thin lines in Fig. \ref{fig:avalanche
probabilities}. Let us focus on the subfigures from $|0,-2\rangle_p$
through $|0,2\rangle_p$. Due to the mixture of $|0,-2\rangle_p$ and
$|0,2\rangle_p$, once SMM evolves to the state $|0,-2\rangle_p$,
there is certain probability that SMM continues evolving to
$|0,2\rangle_p$ directly, without through
$|1,-3/2\rangle_p\rightarrow |0,-1\rangle^-_p\rightarrow
|1,-1/2\rangle^-_p\rightarrow |0,0\rangle_p\rightarrow
|1,1/2\rangle^-_p\rightarrow |1,1\rangle_p\rightarrow
|1,3/2\rangle^-_p\rightarrow |0,2\rangle_p$. As a result, the
probabilities of these intermediate states become smaller compared
with those when $B_2=B_4=0$. Accordingly, the reversal time is also
slightly shortened, as shown by the dashed lines in Fig.
\ref{fig:avalanche}.

The hysteresis loops in the presence of the transverse anisotropy
are shown in Fig. \ref{fig:hysperturb}. The hysteresis loop when
$V_g=-19.6$ is not changed, because it is determined by $E_{|1,\pm
19/2\rangle^-_p}-E_{|0,\pm 10\rangle_p}$, where both $E_{|1,\pm
19/2\rangle^-_p}$ and $E_{|0,\pm 10\rangle_p}$ are hardly affected
by the transverse anisotropy. On the other side, the hysteresis loop
when $V_g=-17.2$ is slightly modified because it is determined by
$E_{|1,\pm 5/2\rangle^-_p}-E_{|0,\pm 2 \rangle_p}$, while $E_{|1,\pm
5/2\rangle^-_p}$ and $E_{|0,2\rangle_p}$ are reshaped by the
transverse anisotropy.

According to Fig. \ref{fig:band perturbed}, the energies of $|0,\pm
10\rangle_p$ and $|1,\pm 21/2\rangle^-_p$ are barely affected by
weak $B_2$ and $B_4$. Therefore, the measurements discussed in Sec.
\ref{sec:measurement} are not affected noticeably. Although the spin
selection rules now allow transitions between $|1, 21/2\rangle^-_p$
and states other that $|0,10\rangle_p$, e.g., $|1, 21/2\rangle^-_p
\leftrightarrow |0,9\rangle_p$, these kind of transitions have
probabilities of order of $(B_{2}/D)^2\sim 10^{-4}$, thus is hard to
be measured.

\section{\label{sec:theta}Non-collinear case}

\begin{figure}[htbp]
\centering
\includegraphics[width=0.5\textwidth]{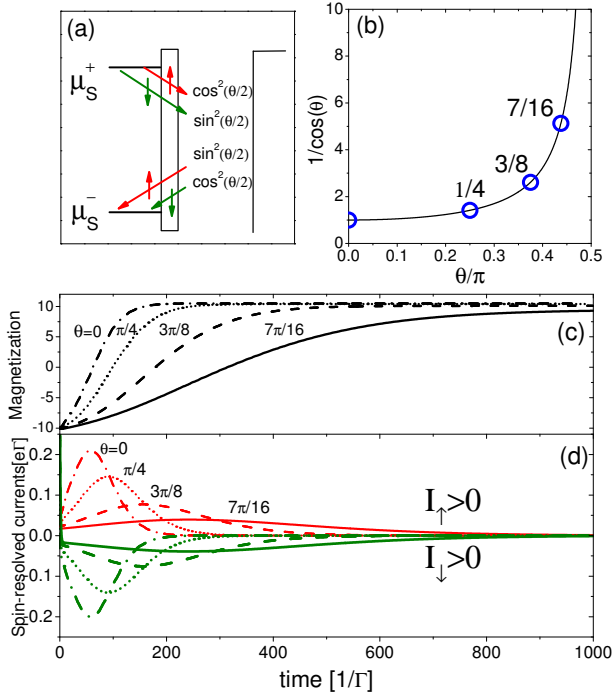}
\caption{(a) In the presence of an angle $\theta$ between the easy
axis of SMM and the spin quantization direction of the lead, there
are four possible injection and leakage processes, marked by the
arrows. $\cos^2(\theta/2)$ or $\sin^2(\theta/2)$ indicates the
relative probability of each process. (b) The reversal time will be
prolonged by $1/\cos(\theta)$. [(c) and (d)] Magnetization and
spin-resolved currents for the same situation shown in Fig.
\ref{fig:avalanche} for different $\theta$. } \label{fig:theta}
\end{figure}

In this section, we will analyze the influence to the three main
results presented in Sec. \ref{sec:numericals}, when considering an
angle $\theta$ between the easy axis of the SMM and the spin
quantization direction of the source lead. Because we have shown
that the weak transverse anisotropy brings ignorable effect in Sec.
\ref{sec:transverse}, we only consider $B_2=B_4=0$ in this section.

\subsection{Reversal dynamics $\theta\neq0$ }

When $\theta\neq0$, the $+$ (-) electrons in the source lead can be
injected into SMM as $\uparrow$($\downarrow$) or
$\uparrow$($\downarrow$) electrons, with probabilities of
$\cos^2\frac{\theta}{2}$ and $\sin^2\frac{\theta}{2}$, respectively.
This will prolong the reversal time. We will illustrate it by
considering the same situation of Sec. \ref{sec:avalanche}. As shown
in Fig. \ref{fig:theta}(a), when $\uparrow$ electrons are injected
from $\mu^{+}_{\mathrm{S}}$ with the tunneling rate $\propto
\cos^2\frac{\theta}{2}$, they are also leaking to
$\mu^-_{\mathrm{S}}$ with the tunneling rate $\propto
\sin^2\frac{\theta}{2}$. From the viewpoint of SMM, the $\uparrow$
electrons are injected from the lead at an equivalent tunneling rate
$\propto \cos^2\frac{\theta}{2}-\sin^2\frac{\theta}{2}=\cos\theta$.
Similarly, an extra ratio $\cos\theta$ is also subjected to the
$\downarrow$ electron leaking to the lead. As a result, the reversal
time will $\propto 1/\cos\theta$, as shown by Figs.
\ref{fig:theta}(c) and (d). As shown by Fig. \ref{fig:theta}(b),
$\theta$ should be at least $ > 0.45\pi $ to increase the reversal
time by one order. In other words, the reversal time will not be
prolonged dramatically unless $\theta$ is very close to $\pi/2$. Of
course, the reversal will fail when $\theta=\pi/2$, i.e., when the
spin quantization direction of the lead is perpendicular to the easy
axis of SMM. Besides, as shown by Fig. \ref{fig:theta}(d), during
the reversal, there is still a pure spin current flowing even in the
presence of $\theta\neq 0$, although $\theta$ will introduce the
extra injection of $\downarrow$ electrons from $\mu^+_{\mathrm{S}}$
and the extra leakage of $\uparrow$ electrons to
$\mu^{-}_{\mathrm{S}}$. This is because these two effects will
cancel with each other since both of them are proportional to
$\sin^2\frac{\theta}{2}$.

\subsection{Non-destructive detection $\theta\neq0$}

\begin{figure}[htbp]
\centering
\includegraphics[width=0.5\textwidth]{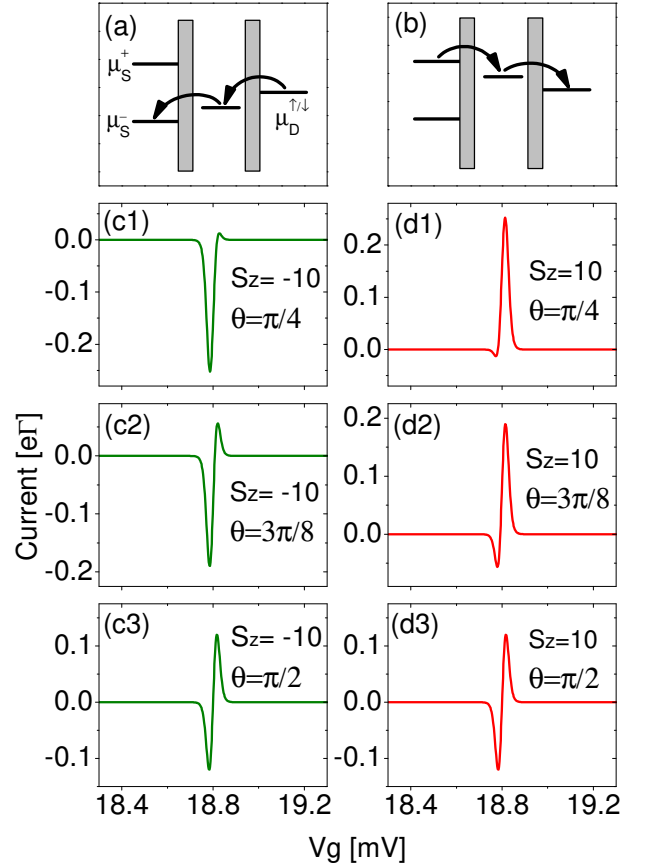}
\caption{(a) Current tends to flow from drain to source when
$E_{|1,\pm21/2\rangle^-}-E_{|0,\pm10\rangle}$ is between
$\mu_{\mathrm{S}}^-$ and $\mu_{\mathrm{D}}^{\uparrow/\downarrow}$.
(b) Current tends to flow from source to drain when
$E_{|1,\pm21/2\rangle^-}-E_{|0,\pm10\rangle}$ is between
$\mu_{\mathrm{S}}^+$ and $\mu_{\mathrm{D}}^{\uparrow/\downarrow}$.
[(c1)-(c3)] The same measurement scheme as Fig.
\ref{fig:measure}(c2) except for $\theta\neq0$. [(d1)-(d3)] The same
measurement scheme as Fig. \ref{fig:measure}(d2) except for
$\theta\neq0$.} \label{fig:theta measure}
\end{figure}

As for the non-destructive measurement in Sec.
\ref{sec:measurement}, the qualitative nature that spin-up (-down)
current is favored when $S_z=S$ ($-S$) is irrelevant to $\theta$.
Besides, our numerical results show that the measurement is still
non-destructive. However, in the presence of $\theta\neq0$, both
$\uparrow$ and $\downarrow$ electrons can tunnel via the Fermi level
$\mu^+_{\mathrm{S}}$ and $\mu^-_{\mathrm{S}}$. As a result, the
current can flow along both directions, different from the
$\theta=0$ results shown in Figs. \ref{fig:measure}(c2) and (d2).

We will use the case $S_z=-S$ to illustrate this difference, as
shown by Fig. \ref{fig:theta measure}(c1)-(c3). When $S_z=-S$, SMM
still favors $\downarrow$ current to tunnel through it via the
transition energy $E_{|1,\pm21/2\rangle^-}-E_{|0,\pm10\rangle}$. As
shown by Fig. \ref{fig:theta measure}(a), when
\begin{equation}\label{DtoS}
    \mu_{\mathrm{S}}^-<E_{|1,\pm21/2\rangle^-}-E_{|0,\pm10\rangle}<\mu_{\mathrm{D}}^{\uparrow/\downarrow}<\mu_{\mathrm{S}}^+,
\end{equation}
current will favor tunneling from drain to source with a relative
probability $\propto \cos^2(\theta/2)$, and this corresponds to the
negative current on the left of Figs. \ref{fig:theta
measure}(c1)-(c3); while as shown by Fig. \ref{fig:theta
measure}(b), when
\begin{equation}\label{StoD}
    \mu_{\mathrm{S}}^-<\mu_{\mathrm{D}}^{\uparrow/\downarrow}<E_{|1,\pm21/2\rangle^-}-E_{|0,\pm10\rangle}<\mu_{\mathrm{S}}^+,
\end{equation}
current will favor tunneling from source to drain with a relative
probability $\propto \sin^2(\theta/2)$, and this corresponds to the
positive current on the right of Figs. \ref{fig:theta
measure}(c1)-(c3).

The case for $S_z=S$ is similar, except that only $\uparrow$ current
is flowing, and the relative ratio between the magnitude of the
negative and positive currents becomes
$\sin^2\frac{\theta}{2}/\cos^2\frac{\theta}{2}$. So one can still
use this difference to distinguish the orientation of the giant
spin, unless $\theta$ is very close to $\pi/2$. When $\theta=\pi/2$,
the negative and positive currents are the same, as shown by Figs.
\ref{fig:theta measure}(c3) and (d3). This is reasonable result,
because when the spin orientation of the source lead is
perpendicular to the easy axis of SMM, one should have symmetric
results for $\uparrow$ and $\downarrow$.

\subsection{Summary for $\theta \neq 0$}

We briefly summarize the influences of $\theta \neq0$. Since
$\theta$ does not change the energy of many-body states, the
steady-state solutions of hysteresis loops shown in Fig.
\ref{fig:hysteresis} are not affected except $\theta\sim\pi/2$. It
will prolong the reversal time by $1/\cos\theta$, as shown in Fig.
\ref{fig:theta}. There are still a pure spin current flowing during
the reversal when $\theta\neq 0$. The non-destructive measurement
shown in Fig. \ref{fig:measure} is still non-destructive when
$\theta\neq0$. However, the measurement signals will be reshaped, as
shown in Fig. \ref{fig:theta measure}. Fortunately, one can still
employ the relative ratio between current flowing along opposite
directions to distinguish the orientation of the giant spin unless
$\theta\sim\pi/2$.

\section{Summary and discussion}

\begin{figure}[htbp]
\centering
\includegraphics[width=0.5\textwidth]{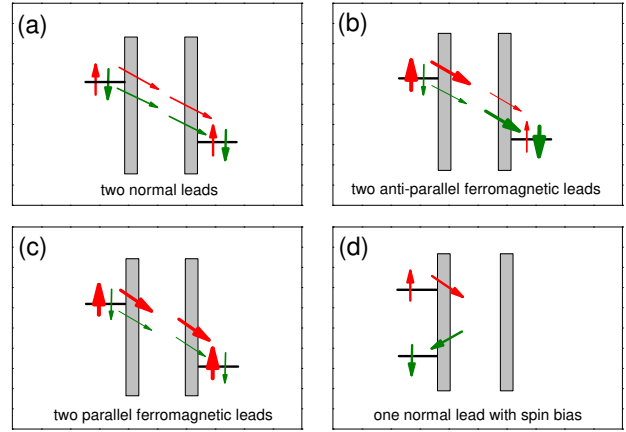}
\caption{[(a)-(c)] Energy schemes employed by the previous authors
and (d) in this work. (a) Nonmagnetic leads (Ref.
\onlinecite{Timm2006prb73235304}), (b) ferromagnetic leads with
anti-parallel polarizations, and (c) ferromagnetic leads with
parallel polarizations (Ref. \onlinecite{Misiorny2007prb76054448});
(d) nonmagnetic lead with spin-dependent splitting of chemical
potentials.} \label{fig:models}
\end{figure}

Before ending this paper, we compare the advantages of the previous
works on current-induced reversal of SMM magnetization to our
proposal. For the setup with one nonmagnetic and one ferromagnetic
lead\cite{Timm2006prb73235304} or the setup with two magnetic leads
that are not fully polarized,\cite{Misiorny2007prb76054448}
electrons injected from one lead can leak to the other lead, while
electrons hopping to one lead can be refilled by electrons from the
other lead, as shown in Figs. \ref{fig:models}(a) and (b). The
leakage and refilling reduces the efficiency of magnetization
reversal because only the excess transmitted spins contribute to the
process of the magnetic reversal.\cite{Timm2006prb73235304} Because
no excess spin is transmitted, two parallel aligned ferromagnetic
leads\cite{Misiorny2007prb76054448} are equivalent to two
nonmagnetic leads when the charge bias is large enough to cover all
the transition energies, as shown in Fig. \ref{fig:models}(c).
Moreover, the electric current induced by the charge bias between
two nonmagnetic electrodes may lead to the decay of SMM
magnetization.\cite{Timm2006prb73235304} In contrast, the current
induced relaxation and low efficiency can be avoided in our one-lead
model, as only a pure spin current flows during the reversal and no
lead-SMM electron exchange is permitted when there is no reversal
occurring. Technically, the ferromagnetic lead can already be
attached to the single molecule with the charge bias voltage easily
exceeding the required threshold voltage,\cite{Pasupathy2004} while
large spin bias over 1 meV still remains an experimental challenge.

In conclusion, the spin bias or spin current can be applied to
control and measure the magnetization of SMM efficiently.\\

\textbf{ACKNOWLEDGEMENTS}

The authors thank Qian Niu, Guoxiang Huang, Ren-Bao Liu, and Rong
L\"{u} for helpful discussions. This work is supported by the
Research Grant Council of Hong Kong under Grant No. HKU 7048/09P and
HKU 10/CRF/08.

\appendix

\section{\label{sec:rateappendix}The Rate equations}

In this appendix, we will present the deduction of the rate
equations Eq. (\ref{rateequation}) following the approach introduced
in Ref. \onlinecite{Blum1996book}. Besides, we will explain why only
diagonal terms of the reduced density matrix are employed for the
present problem. For simplicity, only $\theta=0$ case will be
addressed. The case for $\theta\neq0$ is a straightforward
generalization.

Suppose we have found the eigen states $|n\rangle $ of
$H_{\mathrm{SMM}}$. By using the completeness $\sum_n
|n\rangle\langle n|=$ unity, the Hamiltonian Eqs. (\ref{SMM}) and
(\ref{HT}) can be rewritten as,
\begin{equation}
  H_{\mathrm{SMM}}=\sum_n E_n |n\rangle \langle n |,
\end{equation}
and
\begin{eqnarray}
H_T= \sum_{k,\alpha,\sigma}\sum_{n,m}(V_{k\alpha\sigma}\langle n |
d_{\sigma}|m\rangle c_{k\alpha\sigma}^{\dag}|n\rangle \langle
m|)+h.c..
\end{eqnarray}

The Liouville equation of the density matrix $\hat{\rho}$ of the
entire system is given by
\begin{eqnarray}\label{liouville1}
\partial_t \hat{\rho} = -i[H,\hat{\rho}],
\end{eqnarray}
where $H=H_{\mathrm{SMM}}+H_{\mathrm{lead}}+H_{\mathrm{T}}$. For
arbitrary operator $\hat{O}$, one introduces the interaction picture
$\hat{O}_I(t)=e^{iH_0t}\hat{O}e^{-iH_0t} $, where
$H_0=H_{\mathrm{SMM}}+H_{\mathrm{lead}}$, then Eq.
(\ref{liouville1}) becomes
\begin{eqnarray}\label{liouville2}
\partial_t\hat{\rho}_I(t) =-i[H_{\mathrm{TI}}(t),\hat{\rho}_I(t)
].
\end{eqnarray}
Integrate and iterate Eq. (\ref{liouville2}) for one time
\begin{eqnarray}\label{rho1}
\partial_t\hat{\rho}_I(t)
&\approx& -i[H_{\mathrm{TI}}(t),\hat{\rho}_I(0)]\nonumber\\
&&-\int_0^t dt'[H_{\mathrm{TI}}(t),
[H_{\mathrm{TI}}(t'),\hat{\rho}_I(t')] ].
\end{eqnarray}

The reduced density matrix $\hat{\rho}_{\mathrm{M}}(t)$ of SMM is
obtained from $\hat{\rho}(t)$ by taking the trace over all variables
of the leads, i.e., in the interaction picture,
\begin{eqnarray}
\hat{\rho}_{\mathrm{MI}} \equiv
\mathrm{Tr}_{\mathrm{L}}\hat{\rho}_I.
\end{eqnarray}
Trace out the lead part in Eq. (\ref{rho1})
\begin{eqnarray}\label{rho3}
\partial_t\hat{\rho}_{\mathrm{MI}}(t)
&=&-i\mathrm{Tr}_{\mathrm{L}}[H_{\mathrm{TI}}(t),\hat{\rho}_I(0)]\nonumber\\
&-&\int_0^t dt'\mathrm{Tr}_{\mathrm{L}}[H_{\mathrm{TI}}(t),
[H_{\mathrm{TI}}(t'),\hat{\rho}_I(t')] ].
\end{eqnarray}
The above equation assumed that $H_{\mathrm{T}}$ is switched on at
$t=0$. Prior to this, SMM and the leads are uncorrelated and the
total density matrix is given by their direct products,
\begin{eqnarray}
\hat{\rho}(0) &=& \hat{\rho}_{\mathrm{M}}(0)\otimes
\hat{\rho}_{\mathrm{L}}(0)=\hat{\rho}_{I}(0).
\end{eqnarray}

At this point, we follow Fano to make two key
assumptions.\cite{Fano1957.rmp.29.74} The first is the Born
approximation, which assumes that the leads have so many degrees of
freedom that the effects of interaction with SMM dissipate away
quickly and will not react back to any significant extent, so that
the leads remain described by a thermal equilibrium distribution at
constant temperature $\hat{\rho}_{\mathrm{L}}(0)$ at all time,
\begin{eqnarray}\label{born}
\hat{\rho}_{I}(t) \rightarrow
\hat{\rho}_{\mathrm{MI}}(t)\otimes\hat{\rho}_{\mathrm{L}}(0).\ \ \ \
\mathrm{(Born)}
\end{eqnarray}
The second is the Markoff approximation, which assumes that due to
the rapid relaxation in the leads, correlation functions of lead
electrons decay on a time scale much shorter than the SMM dynamics,
i.e., the leads do not have memory. This allows replacing the
correlation functions by delta-functions in the rates, which are
convolutions of such correlation functions with the reduced density
matrix in the sequential-tunneling approximation. In this context,
$\partial_t \hat{\rho}_{\mathrm{MI}}(t)$ depends only on its present
value $\hat{\rho}_{\mathrm{MI}}(t)$,
\begin{eqnarray}\label{markoff}
\hat{\rho}_{\mathrm{MI}}(t') \rightarrow
\hat{\rho}_{\mathrm{MI}}(t).\ \ \ \ \ \ \ \  \mathrm{(Markoff)}
\end{eqnarray}
Put the two approximations (\ref{born}) and (\ref{markoff}) into Eq.
(\ref{rho3}),
\begin{eqnarray}\label{rho4}
\partial_t\hat{\rho}_{\mathrm{MI}}(t)
&=&-i\mathrm{Tr}_{\mathrm{L}}[H_{\mathrm{TI}}(t),\hat{\rho}_{\mathrm{MI}}(0)\hat{\rho}_{\mathrm{L}}(0)]\nonumber\\
&-&\int_0^t dt'\mathrm{Tr}_{\mathrm{L}}[H_{\mathrm{TI}}(t),
[H_{\mathrm{TI}}(t'),\hat{\rho}_{\mathrm{MI}}(t)\otimes\hat{\rho}_{\mathrm{L}}(0)]
]\nonumber\\
\end{eqnarray}

What follows is straightforward calculation by putting
$H_{\mathrm{TI}}$ into Eq. (\ref{rho4}). The first term on the right
side of Eq. (\ref{rho4}) vanishes because
\begin{eqnarray}
\mathrm{Tr}_{\mathrm{L}}[c_{k\alpha\sigma
I}^{\dag}(t)\hat{\rho}_{\mathrm{L}}(0)]=\mathrm{Tr}_{\mathrm{L}}[c_{k\alpha\sigma
I}(t)\hat{\rho}_{\mathrm{L}}(0)]= 0.
\end{eqnarray}
Besides, the cyclic property of the trace is used
\begin{eqnarray}
\mathrm{Tr}_{\mathrm{L}}[c_{k\alpha\sigma
I}^{\dag}(t)c_{k'\alpha'\sigma'
I}(t')\hat{\rho}_{\mathrm{L}}(0)]=\mathrm{Tr}_{\mathrm{L}}[c_{k\alpha\sigma
I }^{\dag}(t-t') c_{k'\alpha'\sigma'
}\hat{\rho}_{\mathrm{L}}(0)].\nonumber\\
\end{eqnarray}
Finally, one returns to the Schr\"{o}dinger picture, and arrives at
the equation for the arbitrary terms $P_{ij}\equiv\langle
i|\hat{\rho}_M | j\rangle$,
\begin{eqnarray}\label{arbiterms}
&&\partial_t P_{ij}\nonumber\\
&=&-i(E_i-E_j)P_{ij} \nonumber\\
&&-\frac{1}{2}\sum_{i'j'}
\sum_{\alpha\sigma}\Gamma_{\alpha}^{\sigma}\langle i |
d_{\sigma}|j'\rangle \langle j'|d_{\sigma}^{\dag}|i'\rangle
f(E_{j'}-E_{i'}-\mu_{\alpha}^{\sigma}) P_{i'j}\nonumber\\
&&-
\frac{1}{2}\sum_{i'j'}\sum_{\alpha\sigma}\Gamma_{\alpha}^{\sigma}
\langle j' | d_{\sigma}|i'\rangle \langle
i|d_{\sigma}^{\dag}|j'\rangle
[f(E_{i'}-E_{j'}+\mu_{\alpha}^{\sigma})]
P_{i'j} \nonumber\\
&&+\frac{1}{2}\sum_{i'j'}\sum_{\alpha\sigma}\Gamma_{\alpha}^{\sigma}
\langle i | d_{\sigma}|i'\rangle \langle
j'|d_{\sigma}^{\dag}|j\rangle
[f(E_{j'}-E_{j}+\mu_{\alpha}^{\sigma})]
P_{i'j'} \nonumber\\
&&+\frac{1}{2}\sum_{i'j'}\sum_{\alpha\sigma}\Gamma_{\alpha}^{\sigma}\langle
j' | d_{\sigma}|j\rangle\langle i|d_{\sigma}^{\dag}|i'\rangle
f(E_j-E_{j'}-\mu_{\alpha}^{\sigma})P_{i'j'} \nonumber\\
&&+\frac{1}{2}\sum_{i'j'}\sum_{\alpha\sigma}\Gamma_{\alpha}^{\sigma}\langle
i | d_{\sigma}|i'\rangle\langle j'|d_{\sigma}^{\dag}|j\rangle
[f(E_{i'}-E_i+\mu_{\alpha}^{\sigma})]P_{i'j'}\nonumber\\
&&+\frac{1}{2}\sum_{i'j'}\sum_{\alpha\sigma}\Gamma_{\alpha}^{\sigma}\langle
j' | d_{\sigma}|j\rangle\langle i|d_{\sigma}^{\dag}|i'\rangle
f(E_i-E_{i'}-\mu_{\alpha}^{\sigma})P_{i'j'}\nonumber\\
&&-
\frac{1}{2}\sum_{i'j'}\sum_{\alpha\sigma}\Gamma_{\alpha}^{\sigma}\langle
j' | d_{\sigma}|i'\rangle \langle i'|d_{\sigma}^{\dag}|j\rangle
f(E_{i'}-E_{j'}-\mu_{\alpha}^{\sigma}) P_{ij'}\nonumber\\
&&-
\frac{1}{2}\sum_{i'j'}\sum_{\alpha\sigma}\Gamma_{\alpha}^{\sigma}
\langle i' | d_{\sigma}|j\rangle \langle
j'|d_{\sigma}^{\dag}|i'\rangle
[f(E_{j'}-E_{i'}+\mu_{\alpha}^{\sigma})] P_{ij'}\nonumber\\
\end{eqnarray}
The above equation is exactly Eq. (2) of Ref.
\onlinecite{Lehmann2007prl98117203} by replacing $i\rightarrow
\alpha$, $j\rightarrow \beta$, $i'\rightarrow \alpha'$, and $j'
\rightarrow \beta'$.

In this work, we have calculated three physical quantities, the
total magnetization, the LUMO occupation, and the current through
SMM. The general form of $\sigma$ current flowing from lead to SMM
obtained by Lehmann and Loss can be rewritten using our notation
as\cite{Lehmann2007prl98117203}
\begin{eqnarray}\label{lehmanncurrent}
I_{\alpha}^{\sigma}&=&e\mathrm{ Re}
\sum_{ii'j}\Gamma_{\alpha}^{\sigma}
\{f(E_{i}-E_{i'}+\mu_{\alpha}^{\sigma})\langle
j|d_{\sigma}^{\dag}|i' \rangle\langle i'|d_{\sigma}|i \rangle
\nonumber\\
&&\ \ \ \ \ \ \ \ -f(E_{i'}-E_{i}-\mu_{\alpha}^{\sigma})\langle
j|d_{\sigma}|i' \rangle\langle
i'|d_{\sigma}^{\dag}|i \rangle\}P_{ij}.\nonumber\\
\end{eqnarray}
As discussed in Sec. \ref{sec:parameters}, we neglect two
high-energy branches $|1,m\rangle^+$ and $|2,m\rangle$, and consider
only two low-energy branches $|0,m\rangle $ and $|1,m\rangle^-$. In
this context, the terms such as $\langle j | d_{\sigma}|i'\rangle
\langle i'|d_{\sigma}^{\dag}|i\rangle$ must require that $i=j$,
e.g., if $|i'\rangle = |1,m\rangle^- $, only $\langle
0,m-\sigma|d_{\sigma}|1,m\rangle^-$ and $ ^-\langle
1,m|d^{\dag}_{\sigma}|0,m-\sigma\rangle $ are nonzero, so both
$|i\rangle $ and $|j\rangle $ can only be $|0,m-\sigma\rangle$. As a
result, Eq. (\ref{lehmanncurrent}) reduces to
\begin{eqnarray} I_{\alpha}^{\sigma}&=&e
\sum_{ii'}\Gamma_{\alpha}^{\sigma}
\{f(E_{i}-E_{i'}+\mu_{\alpha}^{\sigma})|\langle i'|d_{\sigma}|i
\rangle|^2 \nonumber\\
&&\ \ \ \ \ \ -f(E_{i'}-E_{i}-\mu_{\alpha}^{\sigma})|\langle
i|d_{\sigma}|i' \rangle|^2\}P_{i},
\end{eqnarray} which can be
further simplified as\cite{Timm2006prb73235304}
\begin{eqnarray}
I_{\alpha}^{\sigma} &=& -e\sum_{ij}(n_j-n_i)R_{i\rightarrow
j}^{\alpha\sigma}P_i.
\end{eqnarray}

The LUMO occupation $N$ and the total magnetization $M$ involve only
the diagonal terms of density matrix,
\begin{eqnarray}
N &=& \mathrm{Tr}_{\mathrm{M}}(\hat{\rho}_{\mathrm{M}} \sum_{\sigma}
n_{\sigma})=\sum_{ijk}P_{ij}(\sum_{\sigma}n_{\sigma})_{ki}\nonumber\\
&=&\sum_{i}P_{i}(\sum_{\sigma}n_{\sigma})_{ii},
\end{eqnarray}
and
\begin{eqnarray}
M &=& \mathrm{Tr}_{\mathrm{M}}[\hat{\rho}_{\mathrm{M}}
(S_z+s_z)]=\sum_{ijk}P_{ij}(S_z+s_z)_{ki}\nonumber\\
&=&\sum_{i}P_{i}(S_z+s_z)_{ii},
\end{eqnarray}
because $(s_z+S_z)$ and $\sum_{\sigma}n_{\sigma}$ are eigen
operators of $H_{\mathrm{SMM}}$.

Therefore, one only needs to know the diagonal terms of the density
matrix. By letting $j=i$ in Eq. (\ref{arbiterms}), and employing the
property $\langle i | d_{\sigma}|j'\rangle \langle
j'|d_{\sigma}^{\dag}|i'\rangle\Rightarrow i=i'$, one can readily
show that the equation of $P_i$ only couples to other diagonal terms
of density matrix.

Note that the property $\langle i | d_{\sigma}|j'\rangle \langle
j'|d_{\sigma}^{\dag}|i'\rangle\Rightarrow i=i'$ will be weakly
violated in the presence of the transverse anisotropy $B_2$ and
$B_4$, so that terms such as $ ^-_p\langle
1,m|d^{\dag}_{\sigma}|0,m-\sigma'\rangle_p $ may be also nonzero for
$|\sigma'|>|\sigma|$. However, the relative probability of these
terms\cite{Romeike2006prl96196805} $\sim(B_2/D)^2,\ (B_4/D)^2$, thus
can be omitted for $B_2,B_4\ll D$.

\section{\label{sec:perturbation}Perturbation of Transverse anisotropy}

If we denote $H_{\mathrm{SMM}}=H^{(0)}+H'$, the unperturbed eigen
states of $H^{(0)}$ by $\psi^{(0)}_i$, the first-order correction to
$\psi^{(0)}_i$ is given by
\begin{eqnarray}\label{psi1}
\psi^{(1)}_i &=& \sum_{j\neq i}
\frac{H'_{ji}}{E^{(0)}_i-E^{(0)}_j}\psi^{(0)}_j,
\end{eqnarray}
where $E^{(0)}_i$ is the unperturbed energy of $\psi^{(0)}_i$, and
$H'_{ji}=\langle \psi_j^{(0)}|H'|\psi_i^{(0)}\rangle $. The
second-order (the first order is zero for the present problem)
correction to the energy is given by
\begin{eqnarray}\label{E2}
E_i^{(2)} &=& \sum_{j\neq i}\frac{|H'_{ji}|^2}{E_i^{(0)}-E_j^{(0)}}.
\end{eqnarray}
Finally, the perturbed states and their energies are obtained as
$\psi=\psi^{(0)}+\psi^{(1)}$ and $E=E^{(0)}+E^{(2)}$. Because
$|0,m\rangle$ and $|1,m\rangle^-$ are not coupled by $H'$, we will
discuss them separately.

\subsection{Branch $|0,m\rangle$}

The unperturbed energy is given by
\begin{eqnarray}\label{E00}
E^{(0)}_{|0,m\rangle}= -Dm^2,
\end{eqnarray}
and the perturbation
\begin{widetext}
\begin{eqnarray}\label{H0p}
 && \langle 0,m'|H'|0,m\rangle \nonumber\\
&=&-B_2\sqrt{(S-m)(S+m+1)(S-m-1)(S+m+2)}\ \delta_{m'=m+2}\nonumber\\
&&-B_2\sqrt{(S+m)(S-m+1)(S+m-1)(S-m+2)}\  \delta_{m'=m-2} \nonumber\\
&&-B_4\sqrt{(S-m)(S+m+1)(S-m-1)(S+m+2)}\nonumber\\
&&\ \ \ \times \sqrt{(S-m-2)(S+m+3)(S-m-3)(S+m+4)}\ \delta_{m'=m+4}\nonumber\\
&&-B_4\sqrt{(S+m)(S-m+1)(S+m-1)(S-m+2)}\nonumber\\
&&\ \ \ \times \sqrt{(S+m-2)(S-m+3)(S+m-3)(S-m+4)}\
\delta_{m'=m-4}\nonumber\\
\end{eqnarray}
\end{widetext}
However, because degenerate states $|0,\pm 1\rangle$ and $|0,\pm
2\rangle$ are coupled by $H'$, Eqs. (\ref{psi1}) and (\ref{E2}) can
not be applied directly. One has to perform a linear transformation
first $\Psi^{(0)}_{|0,m\rangle  } = P \psi^{(0)}_{|0,m\rangle }$, so
that
\begin{eqnarray}
\langle
\Psi^{(0)}_{|0,-1\rangle}|H^{(0)}|\Psi^{(0)}_{|0,-1\rangle}\rangle
&\neq& \langle
\Psi^{(0)}_{|0,1\rangle}|H^{(0)}|\Psi^{(0)}_{|0,1\rangle}\rangle\nonumber\\
\langle
\Psi^{(0)}_{|0,-2\rangle}|H^{(0)}|\Psi^{(0)}_{|0,-2\rangle}\rangle
&\neq& \langle
\Psi^{(0)}_{|0,2\rangle}|H^{(0)}|\Psi^{(0)}_{|0,2\rangle}\rangle
\end{eqnarray}
and
\begin{eqnarray} \langle
\Psi^{(0)}_{|0,-1\rangle}|H'|\Psi^{(0)}_{|0,1\rangle}\rangle
=\langle
\Psi^{(0)}_{|0,1\rangle}|H'|\Psi^{(0)}_{|0,-1\rangle}\rangle&=&0\nonumber\\
\langle \Psi^{(0)}_{|0,-2\rangle}|H'|\Psi^{(0)}_{|0,2\rangle}\rangle
=\langle
\Psi^{(0)}_{|0,2\rangle}|H'|\Psi^{(0)}_{|0,-2\rangle}\rangle &=&0.
\end{eqnarray}
Under the same linear transformation, $H^{(0)}$ and $H'$ become
\begin{eqnarray}
\overline{H}^{(0)} = P^{-1}H^{(0)}P,\ \ \overline{H}' = P^{-1}H'P.
\end{eqnarray}
Then one replaces $H^{(0)}\rightarrow \overline{H}^{(0)}$,
$H'\rightarrow \overline{H}'$, and $\psi_i^{(0)}\rightarrow
\Psi_i^{(0)}$ in Eqs. (\ref{psi1}) and (\ref{E2}) to perform the
perturbation.

\subsection{Branch $|1,m\rangle^{\pm}$}

Because $|1,m\rangle^{-}$ and $|1,m\rangle^{+}$ together constitute
the complete set of the subspace of $n=1$, the perturbation of
$|1,m\rangle^-$ has to take $|1,m\rangle^+$ into account. The
unperturbed states and their energy are obtained by diagonalizing
\begin{widetext}
\begin{eqnarray}
&& \left(
  \begin{array}{cc}
  _{\mathrm{LUMO}}\langle m+1/2 |_{\mathrm{GS}} \langle  \downarrow | H^{(0)} |\downarrow\rangle_{\mathrm{LUMO}} |m+1/2\rangle_{\mathrm{GS}} & _{\mathrm{LUMO}}\langle m+1/2 |_{\mathrm{GS}} \langle  \downarrow |H^{(0)}|\uparrow\rangle_{\mathrm{LUMO}}
|m-1/2\rangle_{\mathrm{GS}} \\
   _{\mathrm{LUMO}}\langle m-1/2 |_{\mathrm{GS}} \langle  \uparrow | H^{(0)}|\downarrow\rangle_{\mathrm{LUMO}} |m+1/2\rangle_{\mathrm{GS}}  &_{\mathrm{LUMO}}\langle m-1/2 |_{\mathrm{GS}} \langle  \uparrow | H^{(0)}|\uparrow\rangle_{\mathrm{LUMO}}
|m-1/2\rangle_{\mathrm{GS}} \\
  \end{array}
\right)\nonumber\\
& = & \left(
  \begin{array}{cc}
    \epsilon_0-eV_g+\frac{J}{2}m+\frac{J}{4}-(D+\delta D)(m+\frac{1}{2})^2  &  -\frac{1}{2}J\sqrt{S(S+1)-(m+\frac{1}{2})(m-\frac{1}{2})}  \\
    -\frac{1}{2}J\sqrt{S(S+1)-(m+\frac{1}{2})(m-\frac{1}{2})} & \epsilon_0-eV_g-\frac{J}{2}m+\frac{J}{4}-(D+\delta D)(m-\frac{1}{2})^2  \\
  \end{array}
\right)
\end{eqnarray}
\end{widetext}
for $ m\in \{-S+\frac{1}{2}, S-\frac{1}{2}\} $, so that in each
subspace of $m$,
\begin{eqnarray}
|1,m\rangle^{\pm} &=& \alpha_{m}^{\pm}|\downarrow\rangle
|m+\frac{1}{2}\rangle +\beta_{m}^{\pm }|\uparrow\rangle
|m-\frac{1}{2}\rangle,
\end{eqnarray}
and $\alpha_{|1,S+1/2\rangle}=\beta_{|1,-S-1/2\rangle}=0$ and
$\alpha_{|1,-S-1/2\rangle}=\beta_{|1,S+1/2\rangle}=1$.

\begin{widetext}
The perturbation is then given by
\begin{eqnarray}
&& ^{\nu'}\langle 1,m'|H'|1,m\rangle^{\nu}\nonumber\\
&=&[ (\alpha_{m'}^{\nu'})^*\langle m'+\frac{1}{2}|\langle
\downarrow|+ (\beta_{m'}^{\nu'})^*\langle m'-\frac{1}{2}|\langle
\uparrow|]
H' [\alpha_{m}^{\nu}|\downarrow\rangle |m+\frac{1}{2}\rangle  +\beta_{m}^{\nu }|\uparrow\rangle  |m-\frac{1}{2}\rangle ] \nonumber\\
&=&-\delta_{m'=m+2}B_2[(\alpha_{m'}^{\nu'})^*\alpha_{m}^{\nu}\sqrt{(S-m-\frac{1}{2})(S+m+1\frac{1}{2})(S-m-1\frac{1}{2})(S+m+2\frac{1}{2})}\   \nonumber\\
&&\ \ \ \ \ \ \ \ \ \ \ \ \ \ \ \ \ \ +(\beta_{m'}^{\nu'})^*\beta_{m}^{\nu}\sqrt{(S-m+\frac{1}{2})(S+m+\frac{1}{2})(S-m-\frac{1}{2})(S+m+1\frac{1}{2})} ] \nonumber\\
&&-\delta_{m'=m-2}B_2[(\alpha_{m'}^{\nu'})^*\alpha_{m}^{\nu}\sqrt{(S+m+\frac{1}{2})(S-m+\frac{1}{2})(S+m-\frac{1}{2})(S-m+1\frac{1}{2})}\nonumber\\
&&\ \ \ \ \ \ \ \ \ \ \ \ \ \ \ \ \ \ +(\beta_{m'}^{\nu'})^*\beta_{m}^{\nu}\sqrt{(S+m-\frac{1}{2})(S-m+1\frac{1}{2})(S+m-1\frac{1}{2})(S-m+2\frac{1}{2})}]\nonumber\\
&&-\delta_{m'=m+4}B_4[(\alpha_{m'}^{\nu'})^*\alpha_{m}^{\nu}\sqrt{(S-m-\frac{1}{2})(S+m+1\frac{1}{2})(S-m-1\frac{1}{2})(S+m+2\frac{1}{2})}\nonumber\\
&&\ \ \ \ \ \ \ \ \ \ \ \ \ \ \ \ \ \ \ \ \ \ \ \ \ \times \sqrt{(S-m-2\frac{1}{2})(S+m+3\frac{1}{2})(S-m-3\frac{1}{2})(S+m+4\frac{1}{2})}\nonumber\\
&&\ \ \ \ \ \ \ \ \ \ \ \ \ \ \ \ \  +(\beta_{m'}^{\nu'})^*\beta_{m}^{\nu}\sqrt{(S-m+\frac{1}{2})(S+m+\frac{1}{2})(S-m-\frac{1}{2})(S+m+1\frac{1}{2})}\nonumber\\
&&\ \ \ \ \ \ \ \ \ \ \ \ \ \ \ \ \ \ \ \ \ \ \ \ \
\times \sqrt{(S-m-1\frac{1}{2})(S+m+2\frac{1}{2})(S-m-2\frac{1}{2})(S+m+3\frac{1}{2})}]\nonumber\\
&&-\delta_{m'=m-4}B_4[(\alpha_{m'}^{\nu'})^*\alpha_{m}^{\nu}\sqrt{(S+m+\frac{1}{2})(S-m+\frac{1}{2})(S+m-\frac{1}{2})(S-m+1\frac{1}{2})}\nonumber\\
&&\ \ \ \ \ \ \ \ \ \ \ \ \ \ \ \ \ \ \ \ \ \ \ \ \
\times\sqrt{(S+m-1\frac{1}{2})(S-m+2\frac{1}{2})(S+m-2\frac{1}{2})(S-m+3\frac{1}{2})}\nonumber\\
&&\ \ \ \ \ \ \ \ \ \ \ \ \ \ \ \ \ +(\beta_{m'}^{\nu'})^*\beta_{m}^{\nu}\sqrt{(S+m-\frac{1}{2})(S-m+1\frac{1}{2})(S+m-1\frac{1}{2})(S-m+2\frac{1}{2})}\nonumber\\
&&\ \ \ \ \ \ \ \ \ \ \ \ \ \ \ \ \ \ \ \ \ \ \ \ \
\times\sqrt{(S+m-2\frac{1}{2})(S-m+3\frac{1}{2})(S+m-3\frac{1}{2})(S-m+4\frac{1}{2})}].
\end{eqnarray}
\end{widetext}
Notice that here $m$ are half-integers, so there is not degeneracy
problem. One can employ Eqs. (\ref{psi1}) and (\ref{E2}) directly.



\end{document}